\documentclass[lettersize,journal]{IEEEtran}
\usepackage{amsmath,amsfonts}
\usepackage{algorithmic}
\usepackage{algorithm}
\usepackage{array}
\usepackage{subcaption} 
\usepackage{textcomp}
\usepackage{stfloats}
\usepackage{url}
\usepackage{verbatim}
\usepackage{graphicx}
\usepackage{cite}
\usepackage{makecell}
\usepackage{xcolor}
\begin{document}

\title{Dual-pronged deep learning preprocessing on heterogeneous platforms with CPU, Accelerator and CSD}

\author{Jia Wei, Xingjun Zhang,~\IEEEmembership{Member,~IEEE},Witold Pedrycz,~\IEEEmembership{Fellow,~IEEE}, Longxiang Wang, Jie Zhao
\thanks{This work was supported by the National Natural Science Foundation of China under Grant No. 62372366, 62532005, T2422007, U24A20235 and National Key Research and Development Program of China under Grant No. 2024YFB4505200. (Corresponding author: Xingjun Zhang.)

Jia Wei, is with the Department of Computer Science and Technology, Tsinghua University, Beijing 100084, China and the School of Computer Science and Technology, Xi’an Jiaotong University, Xi’an 710049, China (e-mail: weijia4473@mail.tsinghua.edu.cn).

Xingjun Zhang and Longxiang Wang are with the School of Computer Science and Technology, Xi’an Jiaotong University, Xi’an 710049, China (e-mail: xjzhang@xjtu.edu.cn;wlx419@xjtu.edu.cn).

Witold Pedrycz is with the Department of Electrical and Computer Engineering, University of Alberta, Edmonton, AB T6G 2G6, Canada (e-mail: wpedrycz@ualberta.ca).

Jie Zhao is with the College of Computer Science and Electronic Engineering, Hunan University, Changsha, 410082, China (e-mail: jiezhao@hnu.edu.cn)

}
}



\maketitle

\begin{abstract}
For image-related deep learning tasks, the first step often involves reading data from external storage and performing preprocessing on the CPU. As accelerator speed increases and the number of single compute node accelerators increases, the computing and data transfer capabilities gap between accelerators and CPUs gradually increases. Data reading and preprocessing become progressively the bottleneck of these tasks. Our work, DDLP, addresses the data computing and transfer bottleneck of deep learning preprocessing using Computable Storage Devices (CSDs).  DDLP allows the CPU and CSD to efficiently parallelize preprocessing from both ends of the datasets, respectively. To this end, we propose two adaptive dynamic selection strategies to make DDLP control the accelerator to automatically read data from different sources. The two strategies trade-off between consistency and efficiency. DDLP achieves sufficient computational overlap between CSD data preprocessing and CPU preprocessing, accelerator computation, and accelerator data reading. In addition, DDLP leverages direct storage technology to enable efficient SSD-to-accelerator data transfer. In addition, DDLP reduces the use of expensive CPU and DRAM resources with more energy-efficient CSDs, alleviating preprocessing bottlenecks while significantly reducing power consumption. Extensive experimental results show that DDLP can improve learning speed by up to 23.5\% on ImageNet Dataset while reducing energy consumption by 19.7\% and CPU and DRAM usage by 37.6\%. DDLP also improves the learning speed by up to 27.6\% on the Cifar-10 dataset. 
\end{abstract}


\section{Introduction}
\label{intro}
\IEEEPARstart{I}{n} recent years, Deep Neural Network (DNN) models, represented by Vision Transformer (ViT) \cite{dosovitskiy2020image} and ResNet \cite{he2016deep}, have achieved state-of-the-art performance in computer vision, such as image classification \cite{foret2020sharpness}, object detection \cite{zong2022detrs} and instance segmentation \cite{wang2023internimage}. Benefiting from the great success in the CV field, DNN has also gradually attracted the attention of other fields \cite{jiang2024megascale, zhao2024recommender}.  In particular, the multimodal large model represented by GPT-4o \cite{gpt4o} have triggered a systematic revolution for all over the world.

The success of the models is mainly attributed to the use of large amounts of efficient data for training. On the one hand, in order to accelerate training, new accelerators such as TPUs \cite{you2019fast} and GPUs \cite{choquette2023nvidia} have emerged. On the other hand, the size of deep learning high-performance clusters used for training is keeping growing. The learning process of image-related DNN models is divided into two processes: data preprocessing and training. Data preprocessing involves the CPU reading data from external storage devices (such as NVMe SSDs, SATA SSDs, and HDDs), performing computations, and transferring the preprocessed data to the acceleration device. The training process requires the accelerators to perform forward and backward propagation of the DNN model using the preprocessed data. Therefore, the increase in performance of individual acceleration devices and the growth of high-performance cluster size can only accelerate the training process significantly. Data preprocessing has become a bottleneck for deep learning due to the slow performance improvement of CPUs and storage devices relative to accelerators and cluster size \cite{wei2023fastensor,isenko2022my,nouaji2024speedyloader}.

\begin{figure}[htbp]
  \centering
   \includegraphics[width=0.42\linewidth, trim=255 420 140 120]{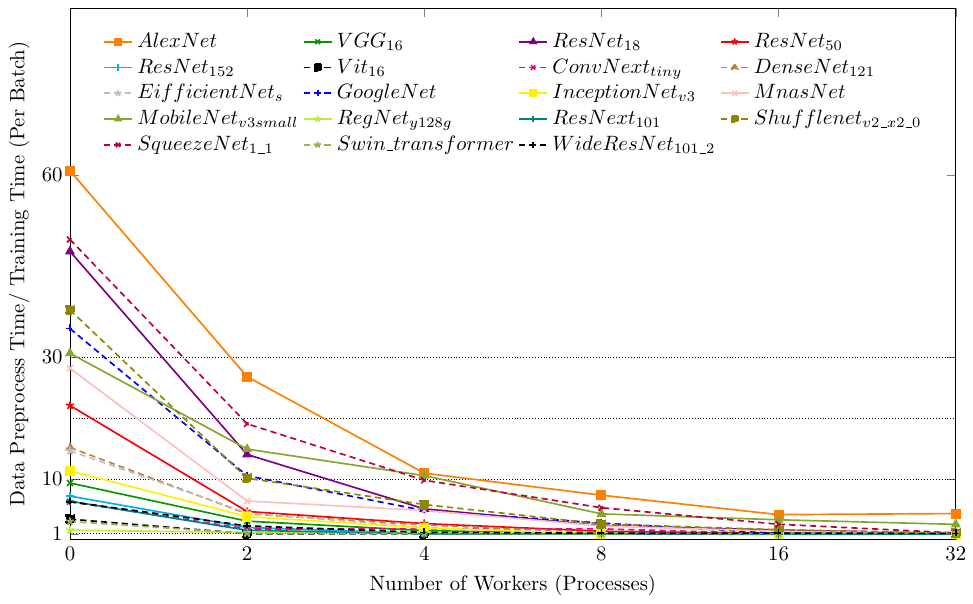}
   \caption{Ratio of Data Preprocessing Time to GPU Training Time vs. Number of Processes. Experiments on 19 Torchvision models with ImageNet show a maximum overhead of 60.67× (mean 20.18×) under single‑process reading. While subprocess parallelism reduces this ratio, it remains above 1 in all configurations, confirming a persistent preprocessing bottleneck. See Section \ref{env} for experimental details.}
   \label{f1}
\end{figure}

As shown in Table \ref{vali}, prior investigations into image-related data preprocessing bottlenecks in deep learning have predominantly targeted the TensorFlow framework and focused on convolutional neural network architectures. In this study, we present the first empirical validation on the PyTorch platform demonstrating that data preprocessing bottlenecks affect both convolutional and transformer-based vision models. This conclusion is corroborated by experiments on 19 different models from the torchvision library using the ImageNet dataset. As shown in Figure \ref{f1}, we observe that the data preprocessing time still exceeds the training time even when loading data using up to 32 subprocesses.
\begin{table}[htbp]
  \centering
  \caption{Validation of Image-Related Deep Learning Data Preprocessing Bottlenecks}
  \label{vali}
  \begin{tabular}{ccccc}
    \hline
    &\textbf{Platform}& \textbf{M-Structure}  & \textbf{M-Number} & \textbf{D-Scale}\\
    \hline
    tf.data \cite{murray2021tf}& Tensorflow  & Conv. &3&140M\\
    HTO \cite{isenko2022my} &Tensorflow& Conv.&3&140M\\
    Pecan \cite{graur2024pecan} &Tensorflow &Conv. &3&140M\\
    Plumber \cite{kuchnik2022plumber} &Tensorflow& Conv. &3&140M\\
    Speedy \cite{nouaji2024speedyloader} &PyTorch & Conv. &1&0.65M\\
    DDLP (ours)&PyTorch & Conv.\& Trans. &19&140M\\
    \hline
  \end{tabular}
\end{table}

In order to improve the efficiency of data preprocessing, some studies have proposed multiprocess and double buffering techniques to accelerate CPU preprocessing \cite{paszke2019pytorch}. However, these methods are still limited by the processing speed of the CPU itself and the data read/write bandwidth of external storage devices. \textbf{How to get rid of the overdependence on CPU, as well as improve the efficiency of the storage side is the key challenge to overcome the data preprocessing bottleneck.}

In recent years, Intelligent Storage Devices (ISDs) such as Computable Storage Devices (CSDs) and Smart SSDs have gradually appeared in various high-performance systems in order to boost the performance of the storage side. Existing CSDs typically use more energy-efficient ARM cores and have shorter I/O paths. CSDs' unique architecture offers the potential to enhance deep learning preprocessing. Recently, there has been some research \cite{heydarigorji2020stannis,do2020cost} exploring CSDs for data preprocessing for simple machine learning tasks. However, these early solutions could only provide simple data manipulation for early machine learning models and datasets but were not available for convolutional or transformer-based deep neural networks. More importantly, all current optimization studies on data preprocessing only consider using a single device (CSD or CPU) to complete data preprocessing operations and fail to overlap data preprocessing with training time effectively. How to efficiently combine CPU and CSD to realize as much computation, I/O overlap and thus faster data preprocessing is the critical challenge we face.

While research on CPU or CSD was developed extensively in the past, their combination was not fully exploited. Hence, we are the first to propose an advanced dual-pronged data preprocessing algorithm (DDLP) that simultaneously uses CPU and CSD to perform data preprocessing operations. 

Designing a two-pronged approach is not an easy task, as it will fundamentally disrupt the traditional data preprocessing process. DDLP faces challenges such as complex task allocation, data distribution, and co-computation with heterogeneous devices. To solve these challenges, we first propose an Moving Towards Each Other (MTE) algorithm, which is pre-given the amount of data that needs to be pre-processed from the CPU and CSD sides based on the processing power of the CPU, the accelerators and the CSD. The MTE algorithm allows the accelerators to process the CSD preprocessed data immediately after using the CPU preprocessed data. MTE enables reliable task and data allocation and efficient CPU and CSD parallel computation.

Although MTE achieves an efficient overlap between CPU preprocessing and CSD preprocessing through a pre-allocation strategy, when the accelerator uses the CSD to preprocess the data, the CSD is left idle, which leads to an additional time overhead problem. In order to further overlap this overhead, we then propose the Weighted Round Robin (WRR) algorithm. It periodically detects whether the CSD has finished preprocessing a batch during the training of the model using classical paths, and if so, uses the CSD preprocessed data. We fully exploit the high read/write bandwidth and independence of the CSD by enabling the CSD and CPU to start data preprocessing from both ends of the dataset at the same time, and at the same time, significantly increase the overlap ratio of preprocessing, CPU preprocessing, and accelerator training.  

DDLP can dynamically select data preprocessed from the CPU or CSD for subsequent training in real-time according to the type of policy, which significantly improves the learning efficiency of deep neural networks in heterogeneous environments, and increases the computational overlap ratio between multiple devices. In addition, DDLP also enables direct data transfer between SSDs and accelerators, which further reduces the consumption of expensive CPU and DRAM resources while improving the transfer efficiency between accelerators and SSDs. 

Finally, we also evaluate the overall training energy consumption after applying DDLP. To validate the effectiveness and stability of DDLP, we experimentally demonstrate the widespread and generalized multiple data preprocessing operations used in training advanced models such as Vit, WRN \cite{Zagoruyko2016}, and ResNet on the most representative ImageNet and Cifar-10 datasets. DDLP can increase the training speed by up to 23.5\% while reducing 19.7\% of the energy consumption and 37.6\% CPU and DRAM utilization. The experiment results show that DDLP can significantly reduce deep neural network training and inference time while decreasing the amount of energy consumed.

Our key contributions can be summarised as follows:
\begin{enumerate}
\item  This work demonstrates through theoretical analysis and a series of detailed experiments that the primary bottleneck in training deep neural networks is predominantly located in the data preprocessing stage.
\item  We propose an advanced dual-pronged data preprocessing algorithm (DDLP) designed to address the challenges of task allocation, data allocation, and co-computation on heterogeneous devices. DDLP achieves training speedup while reducing total training energy consumption. DDLP integrates two adaptive dynamic strategies (MTE and WRR) to achieve efficient computing overlap between CSD data preprocessing, CPU data preprocessing, and accelerator training.
\item The effectiveness and reliability of the DDLP framework are verified through extensive evaluations focusing on training time and energy consumption across various representative models, datasets, data preprocessing operations, and accelerators.
\item Our work has been made open-source, and the publicly accessible website, which will be visible upon acceptance, has been hidden due to the double-blind review policy.
\end{enumerate}

This paper is organized as follows. Section \ref{preliminaries} presents the preliminaries, Section \ref{motivation} introduces the motivation of this paper, Section \ref{method} describes the overall framework of DDLP as well as the MTE and WRR algorithms, Section \ref{imple} gives the details of the DDLP implementation, Section \ref{experiment} presents the experimental results and analyses of DDLP, Section \ref{RW} summarizes related work, and Section \ref{con} summarizes the whole paper and presents the ideas of subsequent research. 

\section{Preliminaries}
\label{preliminaries}
\subsection{Computable Storage Device}
\begin{figure}[H]
    \centering
    \includegraphics[width=0.8\linewidth]{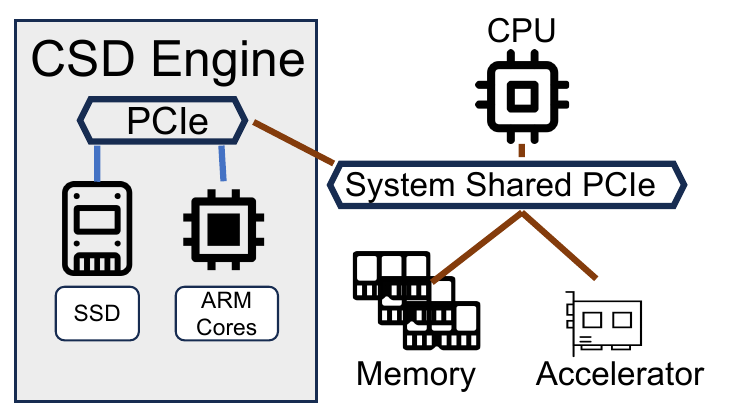}
    \caption{Schematic Diagram of an example environment with CSD.}
    \label{CSD}
\end{figure}

In order to enhance the intelligence and computing power on the storage side, the Computable Storage Device (CSD) is gradually being proposed and applied to high-performance servers. CSD has a higher I/O path and energy-efficient processor compared to host CPU. It can be used directly as an ordinary SSD, but also has the programmable capability itself, which can realize the customized functions on the near-storage side. Its unique architecture is well suited for solving the problems of insufficient computing power and high I/O overhead faced by modern DNN data preprocessing. CSD is a new type of data storage device. CSD reduces the need for data transfer and improves processing efficiency by integrating processing capabilities within or near the storage device, allowing data to be processed or analyzed at the storage location. A typical CSD architecture is shown in Figure \ref{CSD}. 

When data on flash memory are to interact with the host, it must go through the back-end, front-end, and complex NVMe on top of the PCIe link. The long IO paths result in high data transfer overhead for the Host. However, when data interact with the CSD engine, it can bypass the front-end and the power-consuming NVMe link. As a result, the CSD has faster data transfer speeds and lower data transfer overhead compared to the host. Both the host and the CSD have full operating systems and can run applications independently using local processor cores and memory. The Host and CSD interact with each other using tcp/ip. CSDs have the potential to increase data processing speeds, reduce power consumption and costs, and improve system responsiveness. 

\textbf{The unique architecture of the CSD makes it ideally suited to assist the CPU in data preprocessing.} First, the CSD's energy-efficient processor cores are able to perform data preprocessing tasks in parallel with the CPU. Second, CSD's shorter I/O path reduces data transfer overhead.

The shorter data transfer path of CSD makes it well suited for optimizing data-intensive applications. There already exists some results that have demonstrated this claim. For example, HeydariGorji et al. \cite{heydarigorji2020stannis} proposed the Newport CSD architecture and used CSD for a distributed deep learning training task, which improved the acceleration ratio by 2.7x while reducing the energy consumption by 69\%. Cao et al. \cite{cao2020polardb} deployed CSD for the first time to a cloud-native relational database, POLARDB, which reduced the query latency by 30\% and reduced data movement by 50\%. Salamat et al. \cite{salamat2021nascent} proposed a CSD called NASCENT for in-situ database sorting, which improves the speedup ratio by 7.6x and energy efficiency by 5.6x compared to the FPGA baseline. Do et al. \cite{do2020cost} used Newport to perform text compression and similarity detection tasks, improving throughput by 2.42x and energy efficiency by 2.4x compared to the HOST baseline. Despite its greater energy efficiency, CSD is only able to compute at about one-twentieth the speed of the host CPU \cite{heydarigorji2020stannis}. In recent years, some studies have also focused on coordinating the computational and memory resources of CPU and CSD in some very simple tasks. Zhang et al. \cite{zhang2024omnicache} proposed OmniCache to achieve ‘read-CRC-write’ by CPU and CSD together. Yang et al. \cite{yang2023lambda}  proposed $\lambda$\_IO to implement the joint computation of KNN by CSD and CPU.

Although CSD has made initial explorations in the field of Artificial Intelligence (AI) and has implemented cooperation with CPUs on some simple tasks, \textbf{it is still challenging to better leverage CSD to improve the efficiency of deep learning data preprocessing and how to jointly optimize deep learning tasks with CSD and CPU}.

\subsection{Deep Learning}
\label{deepl}
\begin{figure}[htbp]
    \centering
    \includegraphics[width=0.8\linewidth]{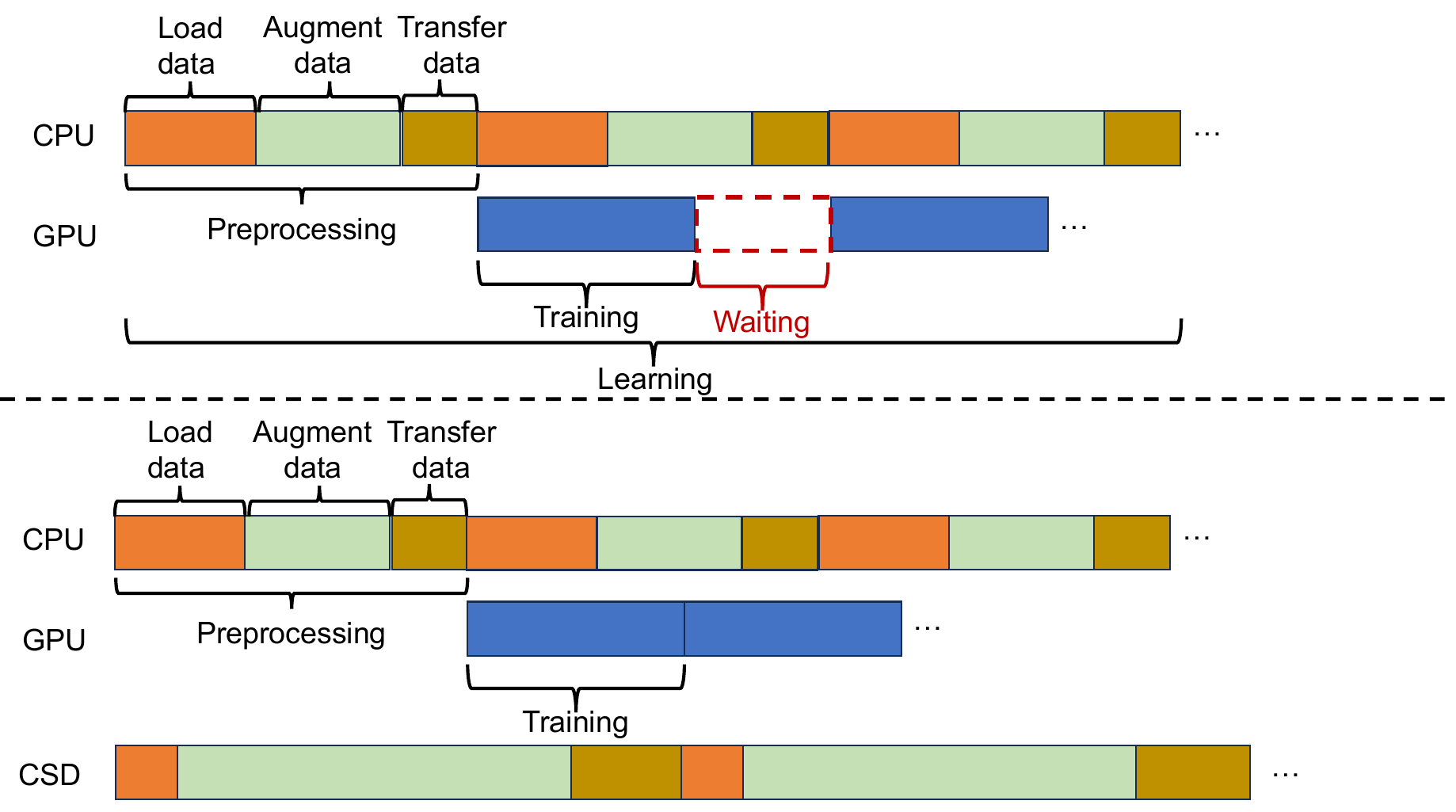}
    \caption{Deep Learning Process. The classical deep learning process (top) and the DDLP deep learning process (bottom).}
    \label{dl}
\end{figure}

Deep learning is the source of current AI superior performance. Effective and efficient deep learning is vital for DNNs. As shown at the top of Figure \ref{dl}, deep learning can be divided into two phases: preprocessing and training. Under the widely used heterogeneous server consisting of CPU and accelerator, the data is first preprocessed by the CPU. The preprocessing phase consists of three parts. The data is first read from the SSD into the DRAM on the CPU side. Then the CPU performs preprocessing operations such as cleaning, rotating, cropping, etc. Finally, the preprocessed data is transferred from DRAM to accelerator memory. In many image preprocessing tasks, the operations involved often consist of sub-tasks that have data dependencies. On one hand, certain sub-tasks must be executed in a specific order to ensure correct results. For example, an image must first be converted into a tensor using $torch.ToTensor()$ before it can undergo normalization with $torch.Normalize()$. On the other hand, the order of sub-tasks can significantly affect both the augmentation outcome and computational efficiency. For instance, applying $torch.RandomResizedCrop(224)$ followed by $torch.RandomHorizontalFlip()$ results in different augmented samples compared to reversing the order of these operations, and the former sequence tends to be more efficient. Effectively managing these dependencies and the order of tasks is essential for optimizing preprocessing pipelines in deep learning workflows. 

In the training phase, the accelerator uses the model and data in memory to train and update the model parameters. The preprocessing and training operations are performed iteratively until the model converges or reaches a predefined termination condition. Due to the imbalance in the development of accelerators and CPUs and storage devices (see Section \ref{motivation} for the detailed reasons), data preprocessing has become the bottleneck of deep learning, and the long waiting time (the red part in Figure \ref{dl}) seriously slows down the deep learning speed, resulting in a high computational and time overhead.

As shown at the bottom of Fig. \ref{dl}, the DDLP proposed in this paper reduces or eliminates waiting overhead by collaborating with the CSD and CPU to complete data preprocessing, thus alleviating the data preprocessing bottleneck. CSD firstly reduces the execution time of the overall data preprocessing phase by sharing the data preprocessing tasks on the CPU side. The accelerator reduces its own waiting time by flexibly selecting and using CPU and CSD preprocessed data. On the other hand, the CSD's own shorter IO paths and more energy-efficient processor cores reduce the cost and energy consumption of data preprocessing. Finally, the reduction of data preprocessing tasks shared by the CPU itself alleviates the need for expensive CPU side compute and DRAM storage capacity and bandwidth.

\section{Motivation}
\label{motivation}
We discussed the data preprocessing bottleneck. In this section, we will directly analyse the causes from multiple perspectives using real data and identify the reasons why they are difficult to mitigate.

\textbf{Computing and transfer performance bottlenecks in preprocessing-side devices.} As we have discussed in Section \ref{deepl}, deep learning includes two phases: data preprocessing and data training or inference. Training and inference of deep learning models are mainly performed on accelerators such as GPUs and DSAs. At the same time, data reading and preprocessing mainly rely on CPUs to read data from external storage devices such as SSDs and then preliminary calculations. Benefiting from new hardware architectures such as tensor cores, accelerators have increased their floating-point computation speed by more than 15x from 2018 to 2023 (from NVIDIA V100 \footnote{V100 GPU.https://www.nvidia.cn/data-center/v100} to NVIDIA H100 \footnote{H100 GPU. https://www.nvidia.cn/data-center/h100}). However, CPU compute speeds have only increased by 2.23x (from Intel core i9 9900K to Intel core i9 13900K \footnote{CPU comparison between 9900k \& 13900k. https://nanoreview.net/en/cpu-compare/intel-core-i9-9900k-vs-intel-core-i9-13900k}), limited by the Moore's law slowdown. Meanwhile, even the most advanced NVMe SSDs are still restricted by PCIe bandwidth, and the growth rate from PCIe 3.0 to PCIe 5.0 \cite{vasa2020pcie} has only been about 4x over the past five years. More importantly, SATA-based SSDs and PCIe 3.0 and 4.0-based SSDs are still the dominant storage devices today. As a result, the growth in performance of devices on the data preprocessing side is far behind the data training and inference side.

\textbf{Limitations on the number of devices on the preprocessing side.} On the other hand, more and more data centers are adopting single-machine multi-accelerator server architectures. For example, the Frontier supercomputing \cite{schneider2022exascale} uses a single-machine quad-accelerator architecture, the Summit supercomputing \cite{yin2019strategies} uses a single-machine 6-accelerator architecture, and the NVIDIA H100 DGX server \cite{choquette2023nvidia} uses a single-machine 8-accelerator architecture. This difference in the number of CPUs and accelerators further exacerbates the development imbalance between the data preprocessing, training, and inference sides. As a result, the data preprocessing side has become a significant bottleneck for deep learning. 

In order to effectively alleviate the preprocessing bottleneck, we need to improve the computational performance of the data preprocessing side, on the one hand, and minimize the data transmission overhead, on the other. The CSD itself is equipped with energy-efficient cores with certain processing capabilities. Meanwhile, the shorter I/O path of CSD has less data transfer overhead. The exploration and utilization of CSD is still in an early stage and has the potential to be further used for deep learning. However, CSD processor cores are more energy-efficient and much less fast than CPUs. Simply using CSD is not enough to face the huge data preprocessing demands. \textbf{How to fully coordinate and utilize CPUs and CSDs, and make CSDs the right hand of CPUs is a key topic with both opportunities and challenges.}

\section{Dual-pronged deep learning preprocessing}
\label{method}
\begin{figure}[htbp]
    \centering
    \includegraphics[width=0.8\linewidth,trim=0 0 0 0]{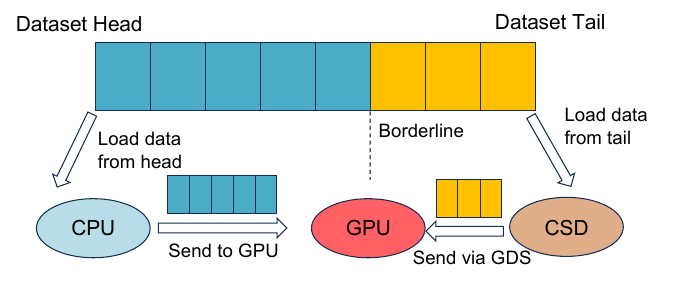}
    \caption{Schematic Diagram of DDLP Architecture}
    \label{f2}
\end{figure}

Table \ref{overlap} provides an overview of the overlap between system computation and communication after applying the DDLP. As shown in Fig. \ref{f2}, the core idea of DDLP is to have the CSD and CPU start parallel preprocessing from both ends of the datasets at the same time. 

\subsection{Challenges}
When the CPU and CSD are employed concurrently for data preprocessing, three key challenges arise:
\begin{enumerate}
\item Determining which device should preprocess each portion of the data so as to minimize total preprocessing time. 
\item Deciding when the accelerator should consume data produced by each preprocessing device for training.
\item Balancing computational efficiency against the required ordering of preprocessing operations. 
\end{enumerate}
To address these challenges, we first introduce a pre-allocation-based MTE method. MTE ensures extensive overlap of computation among the CPU, CSD, and accelerator while preserving data order by having the accelerator process CPU-preprocessed data prior to CSD-preprocessed data. Building on this foundation, we then propose a real-time state detection-based WRR strategy. WRR increases overlap between CSD and accelerator by enabling flexible consumption of CPU- and CSD-processed data, at the expense of partially relaxing strict ordering constraints.

\begin{table}[htbp]
  \centering
  \caption{Computing and Communication Overlap in PyTorch, MTE, and WRR}
  \label{overlap}
  \begin{tabular}{cccc}
    \hline
    \textbf{task}& \textbf{PyTorch}  & \textbf{MTE} & \textbf{WRR}\\
    \hline
    CSD Preprocess& \texttimes  & \checkmark & \checkmark\\
    Transfer CSD Data&\texttimes & \texttimes & \checkmark\\
    CPU Preprocess&\checkmark & \checkmark & \checkmark\\
    Transfer CPU Data&\checkmark & \checkmark & \checkmark\\
    Accelerator Train CPU Data&\checkmark& \checkmark & \checkmark\\
    Accelerator Train CSD Data&\texttimes & \texttimes & \checkmark\\
    \hline
  \end{tabular}
\end{table}

\subsection{MTE Data Flow and Compute Flow Control}
\begin{figure}[htbp]
    \centering
    \includegraphics[width=0.8\linewidth,trim=0 0 0 0]{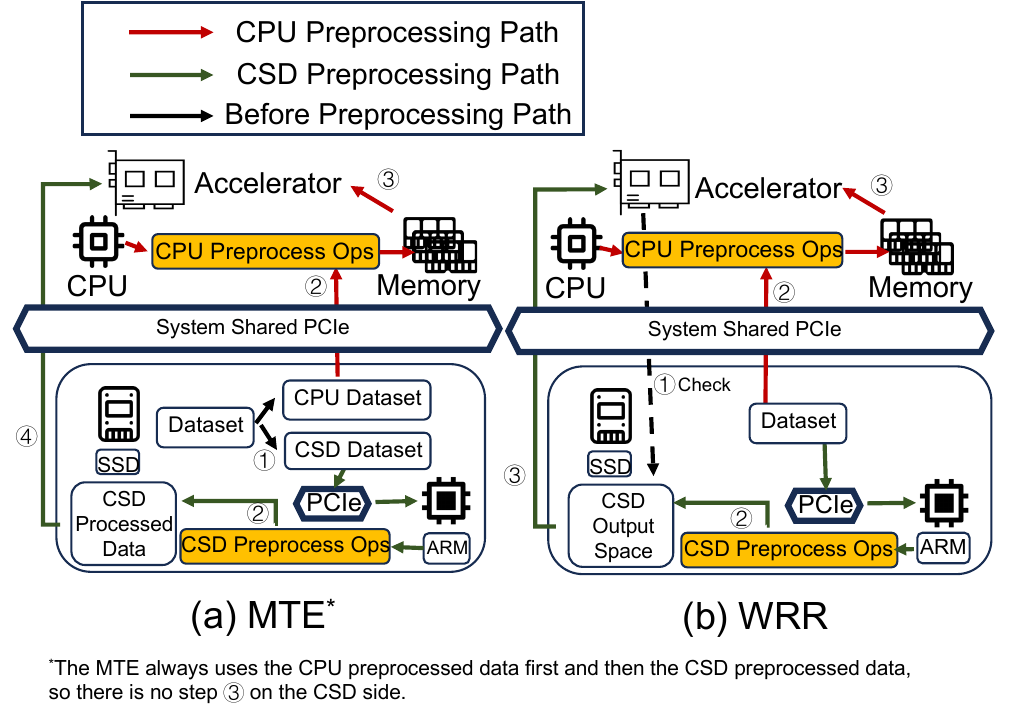}
    \caption{DDLP training process with (a) MTE and (b) WRR.}
    \label{DDLPTP}
\end{figure}

\begin{algorithm}[htbp]
\caption{MTE}
\footnotesize
\label{a1}
\begin{algorithmic}[1]
\REQUIRE Initial model $\mathcal{M}$, number of epochs for training $e$, training dataset $\mathcal{B}$.
\ENSURE The trained model $\mathcal{M}^*$
\STATE $n \gets \text{getsizeof}(\mathcal{B}), i = 0,\ n_{CPU} = n,\ n_{CSD} = 0$
\WHILE{$i < e$}
\IF{$i == 0$}
    \STATE \COMMENT{Determine CPU and CSD Datasets} \\
    \STATE $t_{CPU} =  \text{Teniter\_ppfromheadnCPU}(\mathcal{B})$
    \STATE $t_{CSD} =  \text{Teniter\_ppfrometailonCPU}(\mathcal{B})$
    \STATE $n_{CPU}, \ n_{CSD} = \text{determinenumber}(t_{CPU},t_{CSD})$
    \FOR{$j < n_{CPU}$}
        \STATE $\mathcal{M} = \text{trainmodeluseCPU}(\mathcal{M},\mathbf{B})$
    \ENDFOR   
    \FOR{$j < n_{CSD}$}
        \STATE $\mathcal{M} = \text{trainmodeluseCSD}(\mathcal{M},\mathbf{B})$
    \ENDFOR
\ENDIF
\FOR{$j < n_{CPU}$}
    \STATE $\mathcal{M} = \text{trainmodeluseCPU}(\mathcal{M},\mathbf{B})$
\ENDFOR   
\FOR{$j < n_{CSD}$}
    \STATE $\mathcal{M} = \text{trainmodeluseCSD}(\mathcal{M},\mathbf{B})$
\ENDFOR
\ENDWHILE
\STATE $\mathcal{M^*} =  \mathcal{M}$
\RETURN $\mathcal{M^*}$
\end{algorithmic}
\end{algorithm}

As described in section \ref{intro}, data preprocessing has become a major bottleneck limiting throughput during image-related deep learning. For the traditional preprocessing path, the System Interconnect (PCIe) communication from SSD to Host has a huge overhead. At the same time, because the preprocessing operations are performed by the CPU only, the CPU and DRAM bandwidth consumption is very severe. However, DDLP relocates part of the data preprocessing tasks from the CPU to the CSD, which reduces the total amount of data transferred from the SSD to the Host, and at the same time eases the CPU and DRAM usage. At the same time, since the transfer between the processor and the SSD inside the CSD uses the internal switch, it reduces the usage of the system PCIe and improves the communication speed.

Figure \ref{DDLPTP}(a) illustrates the detailed process of MTE. This process can be used for any image-related deep learning task, such as image classification, object detection, and image-to-text conversion. Unlike traditional data preprocessing pipelines, we must first identify different data flows for different devices. Due to the significant performance difference between CPUs and CSDs, directly equalizing the datasets incurs significant waiting overhead. An intuitive and concise idea is to pre-allocate different numbers of data preprocessing tasks based on relative CPU and CSD performance. First, the accelerator trains the model using data from the CPU side. When the CPU-side data is exhausted, it is coincidental that the CSD finishes its own preprocessing task. The accelerator can then use the preprocessed data directly. Based on this idea, we propose MTE. The MTE training process is shown in Algorithm \ref{a1}. \textcircled{1} At the beginning of model training, we first determine the relative performance $p_{CPU}$ and $p_{CSD}$ of CPU and CSD for different datasets and preprocessing pipelines. This process is done automatically at the start of training and incurs no additional overhead. Specifically, we statistically measure the average time for CPU and CSD to train 10 batches, $t_{CPU}$ and $t_{CSD}$,the processor performance is inversely proportional to the training time, i.e:
\begin{equation}
    \frac{p_{CPU}}{p_{CSD}} = \frac{t_{CSD}}{t_{CPU}}
\end{equation}
Determine the number of CPU- and CSD-trained batches $n_{CPU}$ and $n_{CSD}$ at each epoch based on $p_{CPU}$ and $p_{CSD}$. The $n_{CPU}$ and $n_{CSD}$ are calculated as:
\begin{align}
    n_{CPU} = n \times \frac{p_{CPU}}{p_{CSD}+p_{CPU}} \\
    n_{CSD} = n \times \frac{p_{CSD}}{p_{CSD}+p_{CPU}}
\end{align}
where $n$ is the sample number of the dataset.

Subsequently, MTE implements parallel CPU and CSD preprocessing by controlling the computing flow. \textcircled{2} The CPU performs preprocessing operations from the head of the dataset, batch by batch. Meanwhile, CSD starts to preprocess the data from the tail of the dataset and saves the preprocessed data to the SSD. \textcircled{3} Then the accelerator obtains the CPU preprocessed data for training. This head-to-tail simultaneous processing strategy realizes a high degree of computational overlap between CSD, CPU, and accelerator, reducing the CPU burden under traditional data preprocessing pipelines. \textcircled{4} After the accelerator has trained the $n_{CPU}$ batches of data preprocessed on the CPU side, the accelerator utilizes the GDS technology \cite{daliref} to directly read the CSD preprocessed data from the SSD to the accelerator, and then continues to train the model using this data.

MTE enables overlap between CSD preprocessing and other computations, including CPU preprocessing, CPU to accelerator transfers, and accelerator computations. It is worth noting that since the MTE determines the data processed by the CPU and CSD before training and the data processed by the Accelerator always comes from the CPU first and then from the CSD, the order in which the data is processed is known and determined, which is very important for some deep learning tasks \cite{huang2024mavil}.


\subsection{WRR Data Flow and Compute Flow Control}
While MTE's method of allocating datasets based on the performance of the first several iterations of training is simple and straightforward, modern computer systems are very complex, and changes in various runtime states may result in changes in the relative performance of the CPU and CSD. Such changes can cause the pre-allocated datasets to be unbalanced across CPU and CSD devices, resulting in one side of the CSD or CPU waiting for the other. Meanwhile, most image-related deep learning tasks are not sensitive to the input order of the data, and even incorporate shuffle operations to increase the uncertainty of the datasets \cite{foret2020sharpness,zhao2024recommender}. Therefore, as shown in Fig. \ref{DDLPTP}(b), we further propose that the WRR controls the data flows and computing flows of the CPU and CSD. In contrast to prior dataset segmentation to control the data flows, \textcircled{1} WRR directly detects the presence of pre-trained data from CSD at regular intervals during the training process using CPU pre-trained data. This approach based on real-time state detection avoids the bias caused by performance prediction. This method introduces only minimal I/O overhead. Specifically, we use the command $len(os.listdir(directory\_path)$ to determine if CSD has completed a batch of data preprocessing. This method relies on the file system’s directory table and does not read file contents or access metadata (such as file size, creation time, or permissions), only returning filenames and subdirectory names. Consequently, the I/O overhead is negligible and this cost is included in the preprocessing time reported in our experiments.

Meanwhile, although MTE can effectively overlap most of the computation and I/O overheads of different devices, we find that when the accelerator processes the data on the CSD side, the CSD is idle, indicating that there is still room for further improvement in device utilization. As a result, WRR introduces a more complex control of the computing flows. First, \textcircled{2} the CPU and CSD still perform data preprocessing in parallel. However, the Accelerator no longer sequentially uses data from the CPU side and the CSD side. \textcircled{3} WRR uses the CSD preprocessed data as soon as it detects the presence of CSD preprocessed data, while the CSD is still processing the subsequent data. This unique design allows WRR to further realize the overlap of CSD preprocessing and accelerator training on the basis of WTE.

The WRR training process is shown in the algorithm \ref{a2}, the accelerator takes turns using the CPU for data preprocessing or reading the CSD preprocessed data from the SSD, depending on the CPU and CSD preprocessing speed.

Specifically, before the start of each iteration, it is determined if there are data that the CSD has finished preprocessing. Once the CSD has completed batch preprocessing, the data generated by the CSD are used. We introduce the variable $total$ to determine whether the entire dataset has been used.

WRR not only enables computational overlap in the MTE, but also promotes overlap between CSD preprocessing and other computations and transfers. These computations and transfers include accelerator computations using CSD preprocessed data, and accelerator fetching data directly from the SSD. Due to the inherent instability of server performance, the order of data processed by WRR may be uncertain, and it sacrifices some stability for higher efficiency.

\begin{algorithm}
\footnotesize
\caption{WRR}
\label{a2}
\begin{algorithmic}[1]
\REQUIRE Initial model $\mathcal{M}$, number of epochs for training $e$, training dataset $\mathcal{B}$.
\ENSURE The trained model $\mathcal{M}^*$
\STATE $n \gets \text{getsizeof}(\mathcal{B})$
\STATE $i = 0$
\WHILE{$i < e$}
\STATE \COMMENT{CPU side}
\STATE $j = 0, total = 0$
\FOR{$j < n$}
    \IF{CSD finished one batch}
        \STATE $\mathcal{M} = \text{trainmodeluseCSD}(\mathcal{M}, \mathbf{b^*})$
        \STATE $total$ += 1
    \ENDIF
    \STATE $\mathbf{b} = \text{preprocessonCPU}(\mathcal{B}[j])$
    \STATE $\mathcal{M} = \text{trainmodeluseCPU}(\mathcal{M},\mathbf{b})$
    \STATE $total$ += 1
    \IF{$total == n$}
    \STATE \text{sendsignaltoCSD()}
    \STATE \textbf{Break}
    \ENDIF
\ENDFOR
\STATE \COMMENT{CSD side}
\STATE j = 0
\FOR{$j < n_{CSD}$}
    \IF{getsignalfromCPU() == True}
    \STATE \textbf{Break}
    \ENDIF
    \STATE $\mathbf{b^*} = \text{preprocessandsaveonCSD}(\mathcal{B}[n-j])$
\ENDFOR
\ENDWHILE
\STATE $\mathcal{M^*} =  \mathcal{M}$
\RETURN $\mathcal{M^*}$
\end{algorithmic}
\end{algorithm}

\subsection{Toy Example}
\begin{figure}[H]
  \centering
  \includegraphics[scale=0.5]{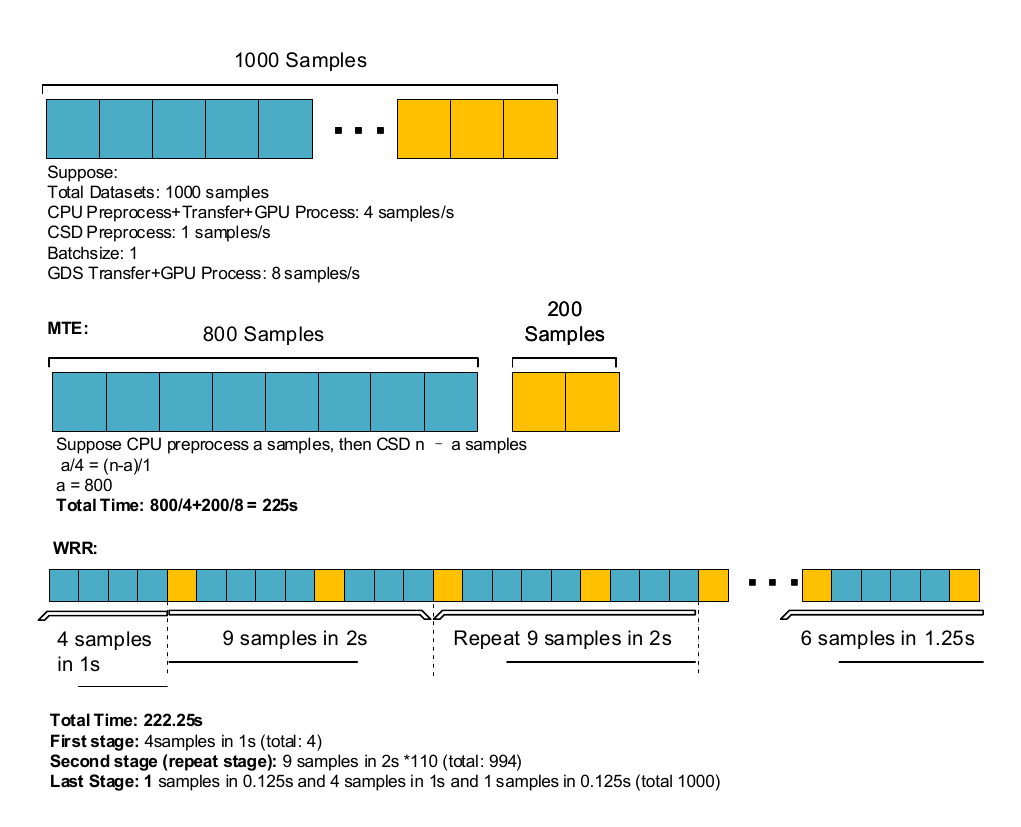}
  \caption{Schematic Comparison of MTE and WRR Algorithms}
  \label{f3}
\end{figure}

Since WRR overlaps more computational overhead than MTE, it achieves a higher acceleration efficiency. As illustrated in Figure \ref{f3}, we assume a dataset of 1000 samples and a batch size of 1. For fairness and consistency with conventional training pipelines, we model the CPU-side preprocessing, data transfer, and accelerator execution as a coupled stage with an effective throughput of 4 samples per second. In contrast, CSD-side preprocessing and SSD storage are modeled at 1 sample per second, while the accelerator reads and processes data via GDS at 8 samples per second.

When using the MTE algorithm, assume that the CPU processes $a$ samples when the CSD and CPU have preprocessed the total 1000 samples together, and $a$ needs to satisfy the following equation \ref{e1}:

\begin{equation}
\label{e1}
\frac{a}{4} = \frac{1000-a}{1}
\end{equation}
The value of $a$ can be calculated as 800 according to equation \ref{e1}, and the total training time can be calculated as 225 seconds according to equation \ref{e2}.
\begin{equation}
\label{e2}
t = \frac{a}{4} + \frac{1000-a}{8}
\end{equation}

Moreover, the total computation time when using the WRR algorithm for the same size dataset can be divided into three phases. The first phase executes the first four batches of CPU-processed batches by the accelerator, which takes a total of 1 second. The second stage cycles through "1 CSD-processed batch, 4 CPU-processed batches, 1 CSD-processed batch, and 3 CPU-processed batches". Each cycle uses 2s to process nine batches and is executed 110 times, taking 220s to process 990 batches. The remaining six batches were executed in the third stage, requiring only 4 CPU-processed batches and 2 CSD-processed batches, for a total of 1.25 s. Thus, it took 222.25 s to train the entire dataset using the WRR algorithm, which is a 1.2\% improvement over MTE.

\subsection{Multi-Accelerator Scenario Scaling}
Previously, we described the design of MTE and WRR in single‑accelerator scenarios. Modern high‑performance servers often pair a single CPU with multiple GPUs, making the extension of DDLP to multi‑accelerator environments both important and challenging. The primary challenge is dispatching CPU‑preprocessed and CSD‑preprocessed data to the correct GPU. To address this, we first leverage PyTorch’s Distributed Data Parallel (DDP) library ($torch.nn.parallel.DistributedDataParallel$) to improve CPU preprocessing. Each GPU runs in its own process, with a dedicated DataLoader for data ingestion and local training. Processes communicate model updates as required. We introduce a $DistributedSampler$ to ensure each process reads a unique partition of the dataset. For load balancing, CSD‑processed data are stored in separate directories keyed by GPU index. In MTE, CSD completes preprocessing for one GPU before switching to the next directory to minimize directory‑switching overhead. In WRR, CSD alternately writes each preprocessed batch across all directories to smooth load distribution.

\section{Implementation}
\label{imple}
This section describes the main software components of DDLP, which is based on the widely used deep learning framework PyTorch \cite{paszke2019pytorch} and the vision library Torchvision \cite{marcel2010torchvision}. We focus on replacing the data preprocessing pathway of PyTorch. DDLP is implemented primarily in Python and Bash, encompassing a total of 1200 lines of code (LOC). The system’s core components are the Host Engine and the CSD Engine. The Host Engine, at approximately 900 LOC, handles CPU-side data preprocessing and orchestrates the MTE and WRR algorithms. The CSD Engine, at roughly 300 LOC, is dedicated to preprocessing on the CSD. For all accelerator targets, the preprocessing pipeline offers an implementation based on the PyTorch DataLoader. In addition, for GPU-based workflows, we provide an alternative implementation using the NVIDIA DALI pipeline (see Section \ref{dali} for details). As such, DDLP can be enabled simply by specifying an option without any compilation operations. This suggests that any image-related deep learning training code implemented using PyTorch can run with DDLP without modification, making DDLP a practical and powerful framework.

\begin{figure}[htbp]
    \centering
    \includegraphics[width=0.8\linewidth]{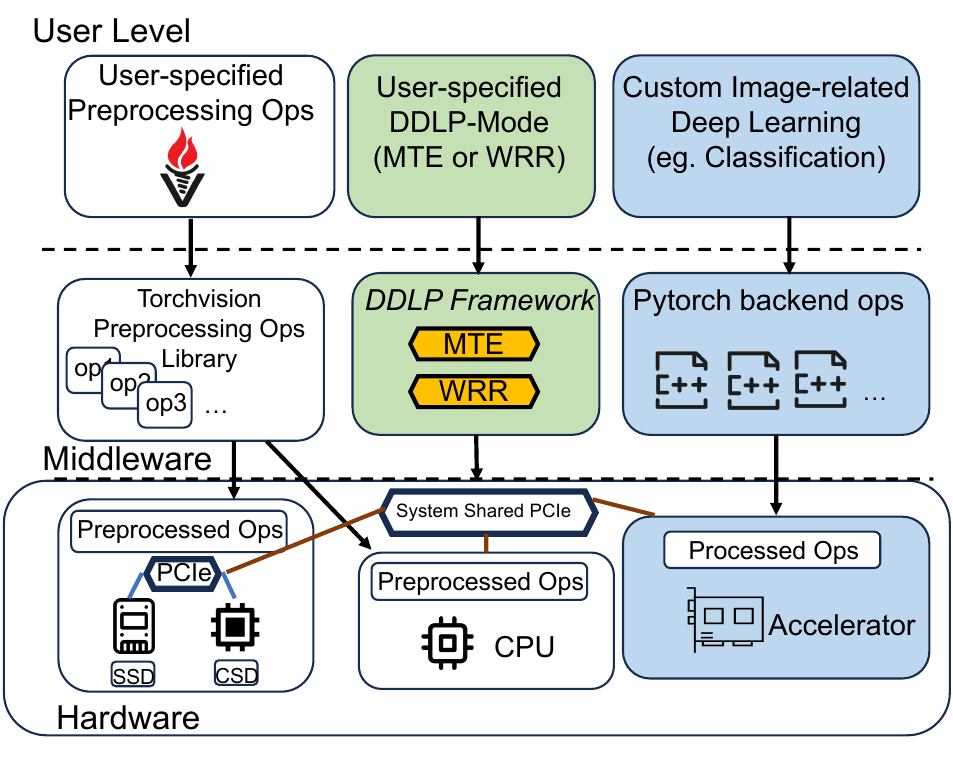}
    \caption{An overview of DDLP framework. Users can combine preprocessing operators, computational tasks and DDLP intrinsic schemes (MTE or WRR) in any way they want.}
    \label{DDLPARC}
\end{figure}

Figure \ref{DDLPARC} \textbf{User Level} shows the design flow for user-defined DDLP. Users can implement a wide variety of custom combinations by specifying the preprocessing operators provided by torchvision, the preprocessing strategies (MTE or WRR) provided by DDLP, and the custom associated deep learning tasks residing in PyTorch. DDLP provides python script templates to implement their own training task combinations. These templates also include a performance analyser and logic checker based on the PyTorch profiler.

Figure \ref{DDLPARC} \textbf{Middleware} describes the interaction between DDLP and the PyTorch and Torchvision runtime engines. PyTorch first calls the high-performance data preprocessing operator library on the backend of torchvision to complete the data preprocessing task, and then uses the obtained results to call the high-performance deep learning operator library on the backend of torch for training. We first carefully modified the data preprocessing path, controlling the data on the SSD to be transferred to the CSD or CPU to perform preprocessing according to different DDLP policies (the preprocessing tasks are identical on different devices). Then, we control the Accelerator to use the preprocessed data on the CPU or CSD side to complete the training according to the specified DDLP policy. The DDLP python script is only used to control the data transfer logic, and the specific preprocessing and training algorithms are derived from high-performance C++/CUDA implementations, so that the overall training still maintains high performance.

Figure \ref{DDLPARC} \textbf{Hardware} shows how the DDLP engine interacts with the hardware components. DDLP uses TCP/IP to implement Host to CSD interaction. During the entire training process, DDLP only needs to send a control signal from Host to CSD once to control CSD to start data preprocessing without any other network communication. It is worth noting that even when WRR is used, the Host only needs to check the length of a given directory to confirm whether it needs to use the pre-processed data from the CSD, without additional hardware communication. When the Accelerator needs to get the pre-processed data from the SSD, it only needs to call the direct storage api to realise the p2p transfer from the SSD to the Accelerator memory.

\section{Experiment and Result}
\label{experiment}
This Section describes the improvement in training speed and energy performance of our data preprocessing using DDLP when training models such as ViT, WRN, and ResNet using the Imagenet and Cifar-10 datasets, respectively. We describe the experimental setting in Section \ref{env}. We first provide a detailed description of the hardware and software configurations of the CPU, accelerators, and CSD. We validate the effectiveness and generalizability of DDLP on two different types of accelerator, GPU and Domain Specific Architecture (DSA). Then the critical hyperparameters of the experimental model, datasets, and preprocessing type are described. We describe the experimental results in Section \ref{res}. We demonstrate that data preprocessing is a bottleneck in deep learning through experiments that train ImageNet using 19 representative models from torchvision. We then present the acceleration performance of DDLP in the ImageNet and Cifar-10 datasets, which are the most representative benchmarks for large and small datasets for image classification tasks, respectively. We finally describe the advantages of DDLP in terms of energy reduction. Finally, we analyze the experimental results in Section \ref{ea}. All experiments in this paper were repeated 20 times using different seeds and averaged as results. The seed may affect the training accuracy, while it has little effect on the learning time and energy consumption, which are the concerns of this paper.

\subsection{Experimental environment}
\label{env}
The hardware and software configurations of the Host (CPU side)  and the CSD side are shown in Table \ref{t2}. Since the current CSD devices are still in the exploratory stage, we use the Pynq platform \cite{viet2021fpga} to emulate the implementation of the CSD devices. The Pynq has similar to the New port CSD \cite{heydarigorji2020stannis} ARM processor, DRAM and external storage. The only difference is that Pynq uses independent external storage instead of interacting directly with Host storage. We simulate implementing the CSD data transfer function by ignoring the transfer time from the Pynq external storage to the Host external storage. Concretely, we first deploy PyTorch and Torchvision on the Pynq platform. We could not deploy the same version of PyTorch and Torchvision for Host and CSD due to operating system versions and processor architectures. However, we kept them as up-to-date as possible and ensured consistent results across platforms. We experimentally confirmed that CSD and CPU preprocess images under the same preprocessing pipeline with the same results even though they use different versions of Torch and Torchvision. When training a model using DDLP, upon receiving a host signal, the Pynq platform iteratively performs the following data preprocessing steps: (1) read a batch of data from the tail of the dataset stored on the SSD; (2) process the data according to the given preprocessing pipeline; and (3) save the processing results to a specified directory on the SSD. We trained Wide ResNet101 (WRN), ResNet152, Vision Transformer (ViT), VGG \cite{Simonyan2015}, and AlexNet \cite{2012ImageNet} on the ImageNet dataset using three different preprocessing pipelines, respectively. The three data preprocessing pipelines are shown in Table \ref{t3}. They are widely used in current ImageNet dataset training by corporate such as NVIDIA \footnote{https://github.com/NVIDIA/aistore/blob/master/docs/tutorials/etl}, Microsoft\footnote {https://github.com/microsoft/nni/blob/master/examples/nas/legacy\\/oneshot/spos/supernet.py} and Xilinx \footnote{https://github.com/Xilinx/VitisAI/blob/master/src/vai\_quantizer\\/vai\_q\_pytorch/example/bfp/resnet/resnet.py}, and academic \cite{liu2022selfsupervised, huang2022green, wei2023fastensor}. The resolution of the images in the ImageNet dataset is uncertain, with the smallest image resolution being $75\times56$, the largest being $4288\times2848$ and the average being $469\times387$. In order to maximize accelerator utilization, the batchsize we use for training is the maximum value that does not incur Out-Of-Memory (OOM) errors. The corresponding batchsizes for different models are shown in Table \ref{t4}. To further validate the performance of DDLP on small datasets, we also used a set of data preprocessing operations \cite{foret2020sharpness}, which currently achieves the best test set accuracy for training from scratch in Cifar-10, to train the Wide ResNet18 model, and the corresponding data preprocessing pipeline and batchsize are shown in Table \ref{t3} and the last row of Table \ref{t4}. The resolution of the images in the Cifar-10 dataset is fixed $32\times32$. Initialization overheads for operations such as context switching and DMA setup have been accounted for in the reported preprocessing times. Through careful configuration of the $num\_workers$ and $batchsize$, these overheads remain negligible relative to the total preprocessing time.

\begin{table}
  \centering
  \caption{Hardware and Software in the Experiments}
  \label{t2}
  \begin{tabular}{cc}
    \hline
    \textbf{Hardware}  & \textbf{Version}\\
    \hline
    CPU & 2*Intel(R) Xeon(R) Silver 4210R\\
    GPU (accelerator 1) & NVIDIA A100 (80GB)\\
    DSA (accelerator 2) & Google TPU (16GB)\\
    NVMe SSD & Samsung 980PRO \\
    CSD & Xilinx Zynq-7000\\
    \hline
    \textbf{Software}  & \textbf{Version}\\
    \hline
    CUDA & 11.7.64 \\
    Host PyTorch & 1.12.1 (GPU) / 2.2.1 (DSA) \\
    Host Torch-xla & 2.2.0 \\
    Host Torchvision &  0.13.1\\
    Host Operating System & Linux Ubuntu 5.15.0-76-generic \\
    CSD PyTorch & 1.7.1 \\
    CSD Torchvision &  0.8.1\\
    CSD Operating System & 4.14.0-xilinx-v2018.3 \\
    \hline
  \end{tabular}
\end{table}

\begin{table}
  \centering
  \caption{Data Preprocessing Pipelines}
  \label{t3}
  \footnotesize
  \begin{tabular}{cc}
    \hline
    \textbf{Name}  & \textbf{Pipeline}\\
    \hline
    ImageNet$_1$ & \makecell[c]{RandomResizedCrop(224)\textrightarrow RandomHorizontalFlip() \\ \textrightarrow ToTensor()\textrightarrow Normalize()}\\
    \hline
    ImageNet$_2$ & \makecell[c]{Resize(256)\textrightarrow
    \textrightarrow CentralCrop(224) \\ \textrightarrow ToTensor()\textrightarrow Normalize()}\\
    \hline
    ImageNet$_3$ &  \makecell[c]{Resize(232),\textrightarrow CentralCrop(224)\\ \textrightarrow ToTensor()\textrightarrow Normalize()}\\
    \hline
    Cifar-10 (GPU) & \makecell[c]{RandomCrop((32,32), 4)\textrightarrow RandomHorizontalFlip()\\ \textrightarrow ToTensor()\textrightarrow Normalize()\textrightarrow Cutout()} \\
    \hline
    Cifar-10 (DSA) & \makecell[c]{RandomResizedCrop(224, scale=(0.05, 1.0))\\\textrightarrow ToTensor()\textrightarrow Normalize()} \\
    \hline
  \end{tabular}
\end{table}

\begin{table}
  \centering
  \caption{Batchsize Used for Model Training}
  \label{t4}
  \begin{tabular}{ccc}
    \hline
    \textbf{Model}  & \textbf{Batchsize} & \textbf{Datset}\\
    \hline
    Wide Resnet101(WRN) \& Resnet152 & 256 & ImageNet\\
    Vision Transformer(ViT) \& VGG & 512 & ImageNet\\
    AlexNet& 4096 & ImageNet\\
    \hline
    Wide Resnet18(WRN18) & 4096 & Cifar-10\\
    Vision Transformer(ViT) & 256 & Cifar-10\\
    \hline
  \end{tabular}
\end{table}

\subsection{Experimental Result}
\label{res}
\subsubsection{Deep Learning Bottleneck Result}
\label{bott}
This Section provides details of the training bottleneck experiment in Figure \ref{f1}. To identify bottlenecks in deep learning, we compared the preprocessing overhead and training overhead of training ImageNet (using ImageNet$_1$ in table \ref{t3} as the preprocessing pipeline) with 19 representative deep learning models from Torchvision, the most widely used deep learning computer vision library. We tested the overall data preprocessing time to training time ratio using the main process (worker number of 0) and multi-process (sub-process number of 2-32, growing in the power of 2) reading data for data preprocessing, respectively. The preprocessing overhead consists of three major components: (1) reading training samples from the SSD into CPU memory, (2) performing data preprocessing on the CPU, and (3) transferring the preprocessed data to the GPU. In contrast, the training overhead encompasses the total time the GPU spends performing forward and backward propagation, as well as updating model parameters. The experimental results are shown in Figure \ref{f1}. The time spent on data preprocessing when using the main process to read data is up to 60.67x the training time, with an average of 20.18x. Although the time spent on data preprocessing generally decreases as the number of subprocesses increases, the ratio of data preprocessing time to training time still exceeds 1 in most cases, even when using up to 32 subprocesses. As the number of sub-processes increases, the occupation of expensive CPU and memory resources increases linearly, and the interference with processes on the host and accelerator becomes severe. In accordance with observations from previous studies \cite{kuchnik2022plumber, graur2024pecan,nouaji2024speedyloader,murray2021tf,isenko2022my}, the experimental results confirm that data preprocessing constitutes a bottleneck in deep learning.

\subsubsection{GPU ImageNet Result}
The results of our experiments in ImageNet are shown in Table \ref{t5}. The experimental results show that both MTE and WRR can improve the training speed in all preprocessing pipelines. WRR slightly outperforms MTE. Compared to CPU preprocessing in a single process, MTE can improve the training speed by up to 21.71\% and 18.67\% on average, and WRR can improve the training speed by up to 23.50\% and 20.19\% on average. Even when using 16 additional processes for data preprocessing, MTE can still improve up to 16.98\% training speed and 7.71\% on average, and WRR can improve up to 17.96\% and 8.48\% on average. Compared to just using CSD for data preprocessing, when the CPU is used for a single process, MTE can increase the training speed by up to 81.31\%, the average increase is 75.72\%, and WRR can increase the training speed by up to 81.37\%, the average increase is 76.39\%. When the CPU was used with 16 additional processes, MTE was able to increase training speed by up to 97.15\%, with an average increase of 86.64\% training speed, and WRR was able to increase training speed by up to 97.15\%, with an average increase of 86.83\% training speed. Furthermore, to validate DDLP in a multi‑accelerator setting, we conducted experiments on a server equipped with two NVIDIA A100 GPUs. We selected ViT and ResNet‑152 to represent transformer‑based and convolutional vision models, respectively. All experiments used the ImageNet dataset with the Imagenet\_1 preprocessing pipeline. As shown in rows 6 and 7 of Table \ref{t5}, DDLP continues to deliver excellent performance. Under MTE and WRR, end‑to‑end throughput improves by 14.51\% and 15.62\% over a single‑threaded CPU baseline, and by 8.96\% and 9.67\% over a 16‑threaded CPU configuration (each GPU 8 threads). Relative to CSD‑only preprocessing, the corresponding speedups are 87.30\% and 87.40\%. These results confirm DDLP’s effectiveness in multi‑GPU scenarios.

\begin{table*}[ht]
  \caption{Average Learning Time (Preprocess Time + Training Time) (s) Per Batch for Different Models with Different Preprocessing Pipelines}
  \centering
  \label{t5}
  \begin{tabular}{ccccccccc}
    \hline
    &\textbf{CPU$_0$}  & \textbf{CPU$_{16}$}& \textbf{CSD}& \textbf{MTE$_0$}& \textbf{WRR$_{0}$}& \textbf{MTE$_{16}$}& \textbf{WRR$_{16}$} &\textbf{Preprocess Type}\\
    \hline
    \textbf{WRN} & 3.527 & 1.779& 10.014& 2.761& \textbf{2.698}& 1.618& \textbf{1.604} &ImageNet$_1$\\
    \textbf{ResNet152} & 3.376 & 1.401& 10.315& 2.672&\textbf{2.624}& 1.308& \textbf{1.301}&ImageNet$_1$\\
    \textbf{ViT} & 8.536 & 7.497& 22.24& 6.996&\textbf{6.695}& 6.388& \textbf{6.171}&ImageNet$_1$\\
    \textbf{VGG}  & 5.522 & 2.418& 19.84& 4.506&\textbf{4.449}& 2.263& \textbf{2.255}&ImageNet$_1$\\
    \textbf{AlexNet}& 48.48 & 5.224& 155.1& 31.24&\textbf{31.12}& 5.111& \textbf{5.104}&ImageNet$_1$\\
    \textbf{ViT (2GPUs)}& 5.428 & 3.765& 22.79& 4.658&\textbf{4.580}& 3.452& \textbf{3.422}&ImageNet$_1$\\
    \textbf{ResNet152 (2GPUs)}& 2.188 & 1.406& 10.08& 1.87&\textbf{1.85}& 1.280& \textbf{1.274}&ImageNet$_1$\\
    \hline
    \textbf{WRN} & 3.572 & 1.748& 12.225& 2.904& \textbf{2.859}& 1.620& \textbf{1.611} &ImageNet$_2$\\
    \textbf{ResNet152} & 3.571 & 1.459& 12.242& 2.883&\textbf{2.845}& 1.369& \textbf{1.364}&ImageNet$_2$\\
    \textbf{ViT} & 9.031 & 7.492& 25.614& 7.458&\textbf{7.198}& 6.513& \textbf{6.351}&ImageNet$_2$\\
    \textbf{VGG}  & 6.001 & 2.460& 23.368& 4.948&\textbf{4.898}& 2.321& \textbf{2.315}&ImageNet$_2$\\
    \textbf{AlexNet}& 42.275 & 6.156& 179.5& 33.54&\textbf{33.43}& 5.111& \textbf{5.109}&ImageNet$_2$\\
    \hline
        \textbf{WRN} & 3.612 & 1.629& 11.30& 2.891& \textbf{2.839}& 1.626& \textbf{1.615} &ImageNet$_3$\\
    \textbf{ResNet152} & 3.558 & 1.5821& 11.65& 2.956&\textbf{2.894}& 1.480& \textbf{1.473}&ImageNet$_3$\\
    \textbf{ViT} & 9.003 & 7.451& 25.74& 7.449&\textbf{7.194}& 6.487& \textbf{6.329}&ImageNet$_3$\\
    \textbf{VGG}  & 5.943 & 2.462& 23.16& 4.906&\textbf{4.857}& 2.323& \textbf{2.316}&ImageNet$_3$\\
    \textbf{AlexNet}& 41.06 & 5.773& 176.4& 33.58&\textbf{33.49}& 5.643& \textbf{5.641}&ImageNet$_3$\\
    \hline
  \end{tabular}
\end{table*}

\subsubsection{GPU Cifar-10 Result}
\begin{figure}[htbp]
\centering
\begin{subfigure}{.23\textwidth}
  \centering
  \includegraphics[scale=0.45,trim=170 540 30 120]{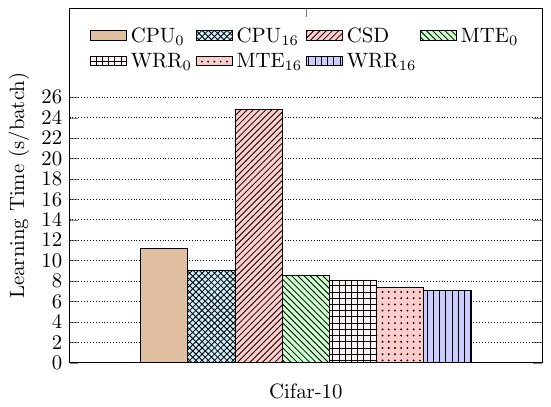}
  \caption{WRN18}
  \label{f4}
\end{subfigure}%
\hfill 
\begin{subfigure}{.23\textwidth}
  \centering
  \includegraphics[scale=0.45,trim=170 540 30 120]{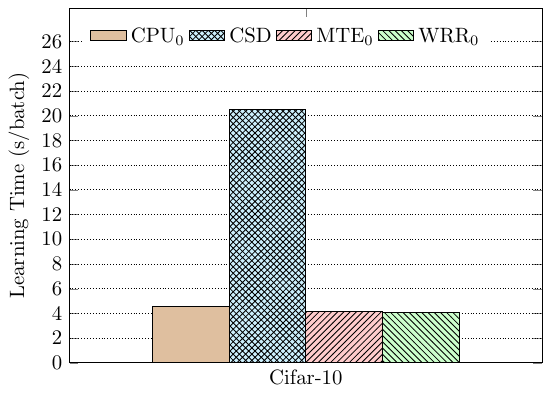}
  \caption{Vit}
  \label{f5}
\end{subfigure}
\caption{Average Learning Time (s) Per Batch for on Cifar-10}
\label{fig:test}
\end{figure}

To further validate the performance of DDLP on small data, we preprocessed and trained the WRN18 model using the Cifar-10 dataset. Figure \ref{f4} shows that similar to the ImagenNet training results, MTE and WRR also achieve consistent training speedup on cifar-10. With a single CPU process, MTE and WRR achieve 23.77\% and 65.59\% and 27.63\% and 67.33\% speedups compared to CPU preprocessing and CSD preprocessing, respectively. With 16 additional CPU processes, MTE and WRR are improved by 18.38\% and 70.20\% and 21.37\% and 71.29\% compared to CPU preprocessing and CSD preprocessing, respectively.


\subsubsection{DSA Cifar-10 Result}
In order to further validate the generalizability of DDLP on different types of acclerator, we preprocessed and trained the ViT model on the DSA platform using the Cifar-10 dataset. The DSA device has smaller memory (16GB) compared to the GPU (80GB) and does not support tuning data reads via $num\_workers$. We used a batchsize of 256 which is the maximum without memory overflow and $num\_workers=0$ for the experiments. Figure \ref{f5} shows that similar to the results of training on GPU, MTE and WRR achieve consistent training speedups on cifar-10. Compared to CPU preprocessing and CSD preprocessing, the speedups of MTE and WRR are 9.70\% and 79.71\%, and 11.13\% and 80.04\%, respectively, when using a single CPU process.

\subsubsection{Co-optimisation with NVIDIA DALI Result}
\label{dali}
NVIDIA DALI \cite{daliref} mitigates the CPU preprocessing bottleneck with optimized preprocessing libraries and transfers a portion of the preprocessing operations to the GPU. However, DALI supports far fewer preprocessing operations than Torchvision and incurs more memory overhead \cite{daliref}. The optimization of data preprocessing by DDLP and DALI is orthogonal. To integrate NVIDIA DALI into DDLP for complete data preprocessing, a pipeline was first instantiated using $nvidia.dali.pipeline.Pipeline$, and all preprocessing operations provided by $torchvision.transforms$ were replaced with equivalent calls to the DALI API. The DALI pipeline was then encapsulated within a PyTorch iterator by employing $nvidia.dali.plugin.pytorch.DALIClassificationIterator$. Finally, this custom iterator supplanted the standard $torch.utils.data.DataLoader$, thereby enabling DDLP to orchestrate and execute the entire data preprocessing workflow through DALI.

As shown in the table \ref{dalic}, we compared the  WRN and Vit models training efficiency of GPU-based integrated DALI for MTE and WRR (MTE\_D \& WRR\_D), GPU- and CPU-based DALI (DALI\_C \& DALI\_G), and torchvision (TV) for using the 16-process $ImageNet_1$ pipeline. MTE\_D and WRR\_D achieve average speedups of 29.78\%, 28.10\%, 5.69\% and 30.29\%, 28.63\%, 6.38\% over TV, DALI\_C, and DALI\_G, respectively.

\begin{table}[ht]
  \caption{Average Learning Time (s) Per Batch for WRN and Vit Models with 16-process $Imagenet_1$ Pipeline.}
  \centering
  \footnotesize
  \label{dalic}
  \begin{tabular}{cccccc}
    \hline
    &\textbf{TV}  & \textbf{DALI\_C}& \textbf{DALI\_G}& \textbf{MTE\_D}& \textbf{WRR\_D}\\
    \hline
    \textbf{WRN} & 1.779 & 1.755& 1.576& 1.460& \textbf{1.450}\\
    \textbf{Vit} & 7.497 & 7.221& 4.558& 4.376& \textbf{4.341}\\
    \hline
  \end{tabular}
\end{table}


\subsubsection{Energy Consumption Result}
DDLP offloads part of the data preprocessing workload from the CPU to a high‐efficiency CSD, enabling both the CPU and CSD to maintain near‐100 \% utilization (comparable to CPU‐only operation) while reducing the total active time required by the CPU and DRAM to complete the preprocessing pipeline. In this Section, we experiment and present the energy-saving performance of DDLP from two perspectives: total DNN learning energy consumption and CPU and DRAM usage time.

\paragraph{Total DNN Learning Energy Consumption Result}
We use two metrics, the average energy consumption of a single batch and the total electricity cost of training 100 epochs, to measure the total energy-saving effect of DDLP. Our energy consumption calculation is simplified as the product of processor power and time. We have 40 threads in the CPU in Table \ref{t2}, with a total power of 200 W. Therefore, we make the power of a single process 5 W and the power of the extra 16 processes 85 W (the main process and the extra 16 processes total 17 processes). The CSD power in Table \ref{t2} is 0.25 W. The average learning energy consumption per batch is shown in Table \ref{t6}. In the single-process experiment, MTE can save up to 17.80\% and average 15.53\% energy consumption, and WRR can save up to 19.68\% and average 17.25\% energy consumption. In additional 16-process experiments, MTE can save up to 14.54\% and average 7.54\% energy consumption, and WRR can save up to 17.44\% and average 8.46\% energy consumption. The experimental results show that MTE and WRR can improve training speed while saving energy consumption. Although using only CSD for preprocessing and training can save more energy consumption, it will significantly increase the total learning time.

Further, we use the Vancouver basic electricity price of \$0.095/kWh to get the total price of electricity consumption under different data preprocessing pipelines, as shown in Table \ref{t6}. The experimental results show that we can save up to \$0.61 and \$0.73 electricity cost by using MTE and WRR for a single ImageNet training, while the daily basic electricity cost of Vancouver households in January 2024 is \$0.21. This result means that our electricity cost savings can satisfy the basic electricity consumption of up to three households in a single day. More importantly, to achieve optimal model performance, the model usually needs to be trained hundreds or thousands of times with different hyperparameters. The electricity cost saved by DDLP also increases super-linearly with the number of training times (since most countries and regions currently charge higher unit prices for more extra electricity consumption). 

\paragraph{CPU and DRAM Usage Time Result}
CPU and DRAM are expensive and scarce resources in servers, and reducing the use of CPU and DRAM can reduce energy consumption while reducing interference with other tasks and improving server efficiency. Since CSD uses its own CPU and DRAM for data preprocessing and uses GDS for data transfer between accelerator and SSD, CSD data preprocessing consumes little of host's CPU and DRAM resources. The experimental results are shown in Table \ref{t8}. In the single-process experiment, MTE can reduce CPU and DRAM resource usage by up to 31.45\% and 25.49\% on average, and WRR can reduce CPU and DRAM resource usage by up to 37.60\% and 27.85\% on average. In the additional 16-process experiment, MTE can reduce CPU and DRAM resource usage by up to 28.72\% and 14.42\% on average, and WRR can reduce CPU and DRAM resource usage by up to 34.34\% and 16.15\% on average. Experimental results demonstrate that MTE and WRR substantially reduce CPU and DRAM usage, thereby lowering cross‐process interference and enhancing server efficiency.

\begin{table*}[ht]
  \caption{Average Learning Energy Consumption (Joule, J) Per Batch / Total Electricity Costs (\$) during 100 Epochs for Different Models}
  \footnotesize
  \centering
  \label{t6}
  \begin{tabular}{cccccccc}
    \hline
    &\textbf{CPU$_0$}  & \textbf{CPU$_{16}$}& \textbf{CSD}& \textbf{MTE$_0$}& \textbf{WRR$_{0}$}& \textbf{MTE$_{16}$}& \textbf{WRR$_{16}$}\\
    \hline
    \textbf{WRN} & 17.63/0.2329 & 151.2/1.997& \textbf{2.504/0.03307}& 14.49/0.1914& \textbf{14.16/0.1871}& 137.9/1.821& \textbf{136.7/1.806}\\
    \textbf{ResNet152} & 16.88/0.2230 & 119.1/1.573& \textbf{2.579/0.03406}& 14.03/0.1852&\textbf{13.77/0.1819}& 111.5/1.472& \textbf{110.9/1.465}\\
    \textbf{ViT} & 42.68/0.2819 & 637.2/4.209& \textbf{5.560/0.03672}& 36.73/0.2426&\textbf{35.15/0.2321}& 544.6/3.597& \textbf{526.1/3.475}\\
    \textbf{VGG} & 27.61/0.1824& 205.5/1.358& \textbf{4.960/0.03276}&23.65/0.1562&\textbf{23.36/0.1543}& 193.0/1.275& \textbf{192.2/1.270}\\
    \textbf{AlexNet}& 192.4/0.1589 & 443.7/0.3665& \textbf{38.77/0.03276}& 164.0/0.1354&\textbf{163.4/0.1349}& 435.7/0.3599& \textbf{435.2/0.3594}\\
    \hline
  \end{tabular}
\end{table*}

\begin{table}
  \caption{Average CPU and DRAM Preprocesing Time (s) Per Batch for Different Models}
  \footnotesize
  \centering
  \label{t8}
  \setlength{\tabcolsep}{5pt}
  \begin{tabular}{cccccccc}
    \hline
    &\textbf{CPU$_0$}  & \textbf{CPU$_{16}$}& \textbf{MTE$_0$}& \textbf{WRR$_{0}$}& \textbf{MTE$_{16}$}& \textbf{WRR$_{16}$}\\
    \hline
    \textbf{WRN} & 2.824 & 1.061 & 2.044& \textbf{1.980}& 0.889& \textbf{0.875}\\
    \textbf{ResNet152} & 2.783 & 0.803& 2.062&\textbf{2.013}& 0.701& \textbf{0.694}\\
    \textbf{ViT} & 5.021 & 3.985& 3.442&\textbf{3.133}& 2.840& \textbf{2.617}\\
    \textbf{VGG} & 4.599& 1.480&3.553&\textbf{3.495}& 1.311& \textbf{1.302}\\
    \textbf{AlexNet}& 37.52 & 4.351& 30.11&\textbf{29.99}& 4.215& \textbf{4.208}\\
    \hline
  \end{tabular}
\end{table}

\subsection{Experimental analyses}
\label{ea}
The experimental results show that DDLP can accelerate preprocessing while reducing the overall learning energy consumption. MTE and WRR can improve performance in all cases, with WRR performing better in most cases.
The acceleration effect of DDLP mainly depends on the following three factors:
\begin{enumerate}
    \item The ratio of CPU-side computation time (CPU-side preprocessing + transfer to accelerator + accelerator computation) to CSD preprocessing time. For MTE and WRR, the CSD preprocessing time is mainly overlapped by the CPU-side computation time. This conclusion indicates that the larger the ratio of CPU-side computation time to CSD preprocessing time, the more CSD preprocessing time overlaps, the higher the acceleration ratio of total learning time.
    \item GDS data reading time. Since the CSD data preprocessing time is overlapped with the CPU side learning time, the additional learning time on the CSD side depends on the GDS data transfer time and the accelerator computation time. The accelerator computation time is the same regardless of whether the data source is CPU or CSD. Therefore, faster GDS data reading time often means faster overall learning time.
    \item The ratio of CSD-side extra learning time (GDS data reading time + accelerator computation time) to CSD-side data preprocessing time. Since WRR can further overlap CSD-side data preprocessing and CSD-side extra learning, the higher the ratio of CSD-side extra learning time to CSD-side data preprocessing time, the more overlapped the CSD preprocessing time and the higher the acceleration ratio of the total learning time. In our experiments, the compute throughput of the CSD is substantially lower than that of the CPU. As a result, the CSD-side extra learning time is small, while the CSD-side data preprocessing time remains comparatively high. This imbalance yields only modest performance gains for WRR over MTE. However, as CSD performance improves, these gains are expected to increase significantly.
\end{enumerate}

The energy-saving effect of DDLP depends on two main factors.
\begin{enumerate}
  \item The number of datasets preprocessed by the CSD. Since the energy consumption generated by the CSD for processing a single batch of datasets is much lower than that of the CPU, the higher the number of datasets that the CSD undertakes to preprocess, the lower the total energy consumed.
  \item The number of CPU processes. Since the relationship between CPU preprocessing speedup and the number of CPU processes is sub-linear and between CPU energy consumption improvement and the number of CPU processes is basically linear, CPU preprocessing using more processes will consume more average energy.
\end{enumerate}
To summarize, CSD is an energy-efficient device capable of reading, writing and preprocessing data with much lower energy consumption. However, CSD's processing speed is much slower than the CPU's. Therefore, it is always an excellent choice to utilize CSD as much as possible when the CSD preprocessing time can be covered by CPU preprocessing time and accelerator training time. When more use of CSD makes its preprocessing time not covered by the CPU and accelerator sides, users must consider the trade-off between total training time and energy consumption.

\section{Related Work and Discussion}
\label{RW}
\subsection{Performance Bottleneck Detection}
Detecting performance bottlenecks is crucial for system optimization. Williams et al. \cite{Williams2009} proposed the Roofline model to identify compute‐ and memory‐bound limits. Ousterhout et al. \cite{ousterhout2017monotasks} and Yu et al. \cite{fetterly2009dryadlinq} applied specialized debugging tools to expose performance issues in big‐data frameworks such as Apache Spark. Naik et al. \cite{naik2016nfvperf} introduced NFVPerf for high‐precision, energy‐efficient bottleneck detection in network function virtualization deployments. In the deep learning domain, several studies on the TensorFlow platform have used trace analysis \cite{kuchnik2022plumber,murray2021tf} and empirical evaluation \cite{graur2024pecan, isenko2022my} to confirm data preprocessing as a bottleneck. Zhao et al. \cite{zhao2022understanding} observed similar bottlenecks in recommendation system models. Nouaji et al. \cite{nouaji2024speedyloader} demonstrated data preprocessing bottlenecks in a 3D image segmentation task using a 3D U-Net model on the PyTorch platform. Nearly all existing work, however, focuses exclusively on convolutional architectures for vision tasks. In contrast, DDLP provides the first experimental validation that both convolutional and transformer-based vision models exhibit data preprocessing bottlenecks. We confirmed this phenomenon in all 19 model architectures in the torchvision library using large-scale ImageNet experiments.

\subsection{Deep Learning Preprocessing Optimization}
To alleviate data preprocessing bottlenecks in deep learning, numerous studies have optimized the preprocessing pipeline from various perspectives. Zhao et al. \cite{zhao2022understanding} designed more efficient data formats. Mohan et al. \cite{mohan2021analyzing} reduced data transfer overhead by caching different data types. Gu et al. \cite{gu2022fluid} improved storage throughput for machine learning–oriented access patterns. Lee et al. \cite{lee2021refurbish} increased throughput by relaxing pipeline consistency guarantees while reducing data randomness. For large‐scale local and remote collaborative training, Um et al. \cite{FastFlow} proposed FastFlow, which uses a smart offloading strategy to assign preprocessing tasks among three candidate locations, using local and remote worker threads. Graur et al. \cite{graur2024pecan} introduced AutoPlacement, which applies resource‐aware scheduling to minimize training overhead. Some systems offload preprocessing tasks onto accelerators—for example, NVIDIA DALI \cite{daliref} targets GPUs, and TrainBox \cite{park2020trainbox} utilizes FPGAs. Notably, DDLP is the first implementation to employ an additional compute‐capable storage device to mitigate data preprocessing bottlenecks. DDLP operates orthogonally to existing optimization schemes, and a combined deployment has the potential to further improve overall performance.

\subsection{Discussion}
\label{sec:discussion}
We decompose one training iteration into $T_{\mathrm{io}}$, $T_{\mathrm{cpu}}$, $T_{\mathrm{csd}}$, and $T_{\mathrm{gpu}}$, denoting the time spent in storage I/O, CPU-side preprocessing, CSD-side preprocessing, and GPU training, respectively. In traditional CPU-only pipelines, the GPU mostly waits for $T_{\mathrm{io}} + T_{\mathrm{cpu}}$, and $T_{\mathrm{gpu}}$ only partially overlaps with data preparation. DDLP changes this balance by offloading part of preprocessing to the CSD, thereby reducing $T_{\mathrm{io}}$ (less data through the host and higher effective I/O throughput) and $T_{\mathrm{cpu}}$ (fewer host operations), while $T_{\mathrm{csd}}$ can overlap with device-level I/O and compute. This decomposition gives a common basis to compare DDLP with GPU-based data loaders and CPU-based reordering schemes and to reason about their potential additive benefits.

GPU-based data loaders such as NVIDIA DALI~\cite{daliref} optimize both the host-side I/O path and CPU preprocessing. Through a pipelined data path and specialized readers, DALI increases the effective $T_{\mathrm{io}}$ throughput from storage to the GPU and reduces $T_{\mathrm{cpu}}$ by migrating a subset of preprocessing operators to the GPU. However, the set of operators efficiently supported on the GPU is more limited than on mature CPU libraries, and GPU-side preprocessing kernels are typically scheduled in sequence with training kernels on the same device, increasing compute and memory-bandwidth load on the GPU (resulting in increasing $T_{\mathrm{gpu}}$) when the model is already close to saturation. In contrast, DDLP reduces $T_{\mathrm{io}}$ and $T_{\mathrm{cpu}}$ by executing operators directly on the CSD and sending only filtered, preprocessed data to the GPU, keeping the GPU focused on training. As we demonstrate in Section~\ref{dali}, DDLP can also be combined with a DALI-optimized data-transfer pipeline. DDLP preprocesses the data stream near storage, and DALI handles the optimized transport to the GPU. This improves data-transfer throughput without adding interference to GPU-side computation, showing that DDLP and DALI are complementary rather than competing mechanisms.

CPU-centric schemes such as PECAN~\cite{graur2024pecan} focus on reordering and batching accesses to improve cache locality and reduce CPU stalls. In our notation, they mainly target a subset of $T_{\mathrm{cpu}}$ by reducing cache misses and improving memory-level parallelism, while leaving $T_{\mathrm{io}}$ unchanged, since data must still be fetched from storage to the host before any reordering takes effect. Consequently, even an ideal PECAN-like scheme that eliminated all CPU-side stalls could only accelerate the portion of the pipeline spent in CPU preprocessing and cannot reduce storage-level I/O wait in I/O-bound regimes. DDLP, by contrast, reduces $T_{\mathrm{io}}$ by moving computation into the CSD and overlapping it with device-side data transfers. Once DDLP has reduced $T_{\mathrm{io}}$ and offloaded part of $T_{\mathrm{cpu}}$ to the CSD, any PECAN-style reordering applied to the remaining CPU operators is bounded by the residual share of $T_{\mathrm{cpu}}$, making PECAN-type optimizations orthogonal to DDLP and, in principle, layerable on top to further optimize the CPU-bound part of the pipeline.

\section{Conclusion, Limitations and Future Work}
\label{con}
In this work, we propose a novel Dual-pronged Deep Learning Preprocessing (DDLP) architecture for CPU and CSD to realize deep learning data preprocessing collaboratively. We propose two algorithms, Moving Towards Each Other (MTE) and Weighted Round Robin (WRR), to accelerate DNN learning while reducing energy consumption by overlapping CSD-side preprocessing with CPU-side preprocessing and accelerator-side computing. We evaluated DDLP from both learning time and total energy consumption perspectives on the large-scale ImageNet dataset and the small-scale Cifar-10 dataset, respectively. We extensively evaluate state-of-the-art WRN, Resnet152, ViT, VGG, AlexNet, and WRN18 models. In ImageNet, our proposed MTE and WRR can reduce 17.80\% and 17.25\% energy consumption while improving the training speed by up to 21.71\% and 23.50\%.  Our proposed MTE and WRR in Cifar-10 can improve the training speed by up to 23.77\% and 27.63\%. We demonstrated that DDLP can further enhance the performance of advanced DALI preprocessing. We also analyze and summarize the reasons for DDLP acceleration and energy savings, providing valuable guidelines to steer subsequent deep learning acceleration and energy savings studies. This paper has limitations in that our study is still designed with a single goal of training speed without considering more diverse user requirements. In some cases, training speed may not be the only objective the user seeks, e.g., the user's aspiration may be the optimal energy consumption in a given time. Therefore, we will further consider CPU and CSD co-preprocessing strategies under given user constraints in future research.

\bibliographystyle{IEEEtran}
\bibliography{aaai24}

@article{liu2022selfsupervised,
  title={Self-supervised learning via maximum entropy coding},
  author={Liu, Xin and Wang, Zhongdao and Li, Ya-Li and Wang, Shengjin},
  journal={Advances in Neural Information Processing Systems},
  volume={35},
  pages={34091--34105},
  year={2022}
}

@article{huang2022green,
  title={Green hierarchical vision transformer for masked image modeling},
  author={Huang, Lang and You, Shan and Zheng, Mingkai and et al.},
  journal={Advances in Neural Information Processing Systems},
  volume={35},
  pages={19997--20010},
  year={2022}
}

@article{wei2023fastensor,
  title={Fastensor: Optimise the Tensor I/O Path from SSD to GPU for Deep Learning Training},
  author={Wei, Jia and Zhang, Xingjun and Wang, Longxiang and Wei, Zheng},
  journal={ACM Trans. on Architecture and Code Optimization},
  volume={20},
  number={4},
  pages={1--25},
  year={2023},
  publisher={ACM New York, NY}
}

@article{schneider2022exascale,
  title={The Exascale Era is Upon Us: The Frontier supercomputer may be the first to reach 1,000,000,000,000,000,000 operations per second},
  author={Schneider, David},
  journal={IEEE Spectrum},
  volume={59},
  number={1},
  pages={34--35},
  year={2022},
  publisher={IEEE}
}

@inproceedings{yin2019strategies,
  title={Strategies to deploy and scale deep learning on the summit supercomputer},
  author={Yin, Junqi and Gahlot, Shubhankar and Laanait, Nouamane and et al.},
  booktitle={2019 IEEE/ACM 3rd Workshop on Deep Learning on Supercomputers},
  pages={84--94},
  year={2019}
}

@article{choquette2023nvidia,
  title={Nvidia hopper h100 gpu: Scaling performance},
  author={Choquette, Jack},
  journal={IEEE Micro},
  year={2023},
  publisher={IEEE}
}

@inproceedings{he2016deep,
  title={Deep residual learning for image recognition},
  author={He, Kaiming and Zhang, Xiangyu and Ren, Shaoqing and Sun, Jian},
  booktitle={Proc. of the IEEE Conf. on computer vision and pattern recognition},
  pages={770--778},
  year={2016}
}

@inproceedings{dosovitskiy2020image,
  title={An Image is Worth 16x16 Words: Transformers for Image Recognition at Scale},
  author={Dosovitskiy, Alexey and Beyer, Lucas and et al.},
  booktitle={Int. Conf. on Learning Representations},
  year={2020}
}

@inproceedings{foret2020sharpness,
  title={Sharpness-aware Minimization for Efficiently Improving Generalization},
  author={Foret, Pierre and Kleiner, Ariel and Mobahi, Hossein and Neyshabur, Behnam},
  booktitle={Int. Conf. on Learning Representations},
  year={2020}
}

@article{zong2022detrs,
  title={Detrs with collaborative hybrid assignments training},
  author={Zong, Zhuofan and Song, Guanglu and Liu, Yu},
  journal={arXiv preprint arXiv:2211.12860},
  year={2022}
}

@inproceedings{wang2023internimage,
  title={Internimage: Exploring large-scale vision foundation models with deformable convolutions},
  author={Wang, Wenhai and Dai, Jifeng and Chen, Zhe and et al.},
  booktitle={Proc. of the IEEE/CVF Conf. on computer vision and pattern recognition},
  pages={14408--14419},
  year={2023}
}

@InProceedings{Zagoruyko2016,
  author    = {S. Zagoruyko and N. Komodakis},
  booktitle = {Proc. of the British Machine Vision Conf.},
  title     = {Wide Residual Networks},
  year      = {2016},
  editor    = {Richard C. Wilson and Edwin R. Hancock and William A. P. Smith},
  publisher = {{BMVA} Press},
  bibsource = {dblp computer science bibliography, https://dblp.org},
}

@Article{2012ImageNet,
  author  = {Krizhevsky, Alex and Sutskever, I. and Hinton, G.},
  journal = {Advances in neural information processing systems},
  title   = {ImageNet Classification with Deep Convolutional Neural Networks},
  year    = {2012},
  number  = {2},
  volume  = {25},
}

@inproceedings{heydarigorji2020stannis,
  title={Stannis: low-power acceleration of dnn training using computational storage devices},
  author={HeydariGorji, Ali and Torabzadehkashi, Mahdi and Rezaei, Siavash and et al.},
  booktitle={57th ACM/IEEE Design Automation Conf.},
  pages={1--6},
  year={2020},
  organization={IEEE}
}

@InProceedings{Simonyan2015,
  author    = {K. Simonyan and A. Zisserman},
  booktitle = {3rd Int. Conf. on Learning Representations},
  title     = {Very Deep Convolutional Networks for Large-Scale Image Recognition},
  year      = {2015},
  editor    = {Yoshua Bengio and Yann LeCun},
}

@article{paszke2019pytorch,
  title={Pytorch: An imperative style, high-performance deep learning library},
  author={Paszke, Adam and Gross, Sam and Massa, Francisco and et al.},
  journal={Advances in neural information processing systems},
  volume={32},
  year={2019}
}

@inproceedings{marcel2010torchvision,
  title={Torchvision the machine-vision package of torch},
  author={Marcel, S{\'e}bastien and Rodriguez, Yann},
  booktitle={Proc. of the 18th ACM Int. Conf. on Multimedia},
  pages={1485--1488},
  year={2010}
}

@article{do2020cost,
  title={Cost-effective, energy-efficient, and scalable storage computing for large-scale AI applications},
  author={Do, Jaeyoung and Ferreira, Victor C and Bobarshad, Hossein and et al.},
  journal={ACM Trans. on Storage},
  volume={16},
  number={4},
  pages={1--37},
  year={2020},
  publisher={ACM New York, NY, USA}
}

@article{viet2021fpga,
  title={FPGA-based Acceleration for Convolutional Neural Networks on PYNQ-Z2.},
  author={Huynh, Thang Viet},
  journal={Int. Journal of Computing and Digital Systems},
  volume={11},
  number={1},
  pages={441--450},
  year={2022},
  publisher={University of Bahrain, Deanship of Graduate Studies and Scientific Research}
}

@inproceedings{vasa2020pcie,
  title={Pcie gen-5 design challenges of high-speed servers},
  author={Vasa, Mallikarjun and Liao, Chun-Lin and Kumar, Sanjay and et al.},
  booktitle={IEEE 29th Conf. on Electrical Performance of Electronic Packaging and Systems},
  pages={1--3},
  year={2020},
  organization={IEEE}
}

@article{you2019fast,
  title={Fast deep neural network training on distributed systems and cloud TPUs},
  author={You, Yang and Zhang, Zhao and Hsieh, Cho-Jui and et al.},
  journal={IEEE Trans. on Parallel and Distributed Systems},
  volume={30},
  number={11},
  pages={2449--2462},
  year={2019},
  publisher={IEEE}
}

@inproceedings{cao2020polardb,
  title={$\{$POLARDB$\}$ meets computational storage: Efficiently support analytical workloads in $\{$Cloud-Native$\}$ relational database},
  author={Cao, Wei and Liu, Yang and Cheng, Zhushi and et al.},
  booktitle={18th USENIX Conf. on file and storage technologies},
  pages={29--41},
  year={2020}
}

@inproceedings{salamat2021nascent,
  title={NASCENT: Near-storage acceleration of database sort on SmartSSD},
  author={Salamat, Sahand and H. Aboutalebi, Armin and Khaleghi, Behnam and et al.},
  booktitle={The 29th ACM Int. Symp. on Field-Programmable Gate Arrays},
  pages={262--272},
  year={2021}
}

@inproceedings{jiang2024megascale,
  title={$\{$MegaScale$\}$: Scaling Large Language Model Training to More Than 10,000 $\{$GPUs$\}$},
  author={Jiang, Ziheng and Lin, Haibin and Zhong, Yinmin and et al.},
  booktitle={Proc. of the 21st USENIX Symp. on Networked Systems Design and Implementation},
  pages={745--760},
  year={2024}
}

@article{zhao2024recommender,
  title={Recommender systems in the era of large language models (llms)},
  author={Zhao, Zihuai and Fan, Wenqi and Li, Jiatong and et al.},
  journal={IEEE Trans. on Knowledge and Data Engineering},
  year={2024},
  publisher={IEEE}
}

@inproceedings{gpt4o,
  title={Langcoop: Collaborative driving with language},
  author={Gao, Xiangbo and Wu, Yuheng and Wang, Rujia and et al.},
  booktitle={Proceedings of the Computer Vision and Pattern Recognition Conference},
  pages={4226--4237},
  year={2025}
}

@inproceedings{isenko2022my,
  title={Where is my training bottleneck? hidden trade-offs in deep learning preprocessing pipelines},
  author={Isenko, Alexander and Mayer, Ruben and Jedele, Jeffrey and et al.},
  booktitle={Proc. of the 2022 Int. Conf. on Management of Data},
  pages={1825--1839},
  year={2022}
}

@inproceedings{nouaji2024speedyloader,
  title={SpeedyLoader: Efficient Pipelining of Data Preprocessing and Machine Learning Training},
  author={Nouaji, Rahma and Bitchebe, Stella and Balmau, Oana},
  booktitle={Proc. of the 4th Workshop on Machine Learning and Systems},
  pages={65--72},
  year={2024}
}

@inproceedings{zhang2024omnicache,
  title={$\{$OmniCache$\}$: Collaborative Caching for Near-storage Accelerators},
  author={Zhang, Jian and Ren, Yujie and Nguyen, Marie and et al.},
  booktitle={22nd USENIX Conf. on File and Storage Technologies},
  pages={35--50},
  year={2024}
}

@inproceedings{yang2023lambda,
  title={$\{$$\lambda$-IO$\}$: A Unified $\{$IO$\}$ Stack for Computational Storage},
  author={Yang, Zhe and Lu, Youyou and Liao, Xiaojian and et al.},
  booktitle={21st USENIX Conf. on File and Storage Technologies},
  pages={347--362},
  year={2023}
}

@article{huang2024mavil,
  title={Mavil: Masked audio-video learners},
  author={Huang, Po-Yao and Sharma, Vasu and Xu, Hu and Ryali, Chaitanya and Li, Yanghao and Li, Shang-Wen and Ghosh, Gargi and Malik, Jitendra and Feichtenhofer, Christoph and others},
  journal={Advances in Neural Information Processing Systems},
  volume={36},
  year={2024}
}

@misc{daliref,
  author = {NVIDIA},
  title = {Pieter Luitjens},
  howpublished = {\url{https://www.private-ai.com/en/2020/01/21/nvidia-dali-speeding-up-pytorch/}},
  year = {2024},
  note = {2024-07-29}
}

@article{kuchnik2022plumber,
  title={Plumber: Diagnosing and removing performance bottlenecks in machine learning data pipelines},
  author={Kuchnik, Michael and Klimovic, Ana and Simsa, Jiri and et al.},
  journal={Proc. of Machine Learning and Systems},
  pages={33--51},
  year={2022}
}

@inproceedings{graur2024pecan,
  title={Pecan:Cost-Efficient ML Data Preprocessing with Automatic Transformation Ordering and Hybrid Placement},
  author={Graur, Dan and Mraz, Oto and Li, Muyu and et al.},
  booktitle={2024 USENIX Annual Technical Conf.},
  pages={649--665},
  year={2024}
}

@article{murray2021tf,
  title={tf. data: A Machine Learning Data Processing Framework},
  author={Murray, Derek G and {\v{S}}im{\v{s}}a, Ji{\v{r}}{\'\i} and Klimovic, Ana and et al.},
  journal={Proc. of the VLDB Endowment},
  volume={14},
  number={12},
  pages={2945--2958},
  year={2021},
  publisher={Association for Computing Machinery}
}

@inproceedings{ousterhout2017monotasks,
  title={Monotasks: Architecting for performance clarity in data analytics frameworks},
  author={Ousterhout, Kay and Canel, Christopher and Ratnasamy, Sylvia and Shenker, Scott},
  booktitle={Proc. of the 26th Symp. on Operating Systems Principles},
  pages={184--200},
  year={2017}
}

@article{fetterly2009dryadlinq,
  title={DryadLINQ: A system for general-purpose distributed data-parallel computing using a high-level language},
  author={Fetterly, Yuan Yu Michael Isard Dennis and Budiu, Mihai and Erlingsson, {\'U}lfar and et al.},
  journal={Proc. LSDS-IR},
  volume={8},
  year={2009}
}

@inproceedings{naik2016nfvperf,
  title={NFVPerf: Online performance monitoring and bottleneck detection for NFV},
  author={Naik, Priyanka and Shaw, Dilip Kumar and Vutukuru, Mythili},
  booktitle={2016 IEEE Conf. on Network Function Virtualization and Software Defined Networks},
  pages={154--160},
  year={2016},
  organization={IEEE}
}

@inproceedings{zhao2022understanding,
  title={Understanding data storage and ingestion for large-scale deep recommendation model training: Industrial product},
  author={Zhao, Mark and Agarwal, Niket and Basant, Aarti and et al.},
  booktitle={Proc. of the 49th annual Int. Symp. on computer architecture},
  pages={1042--1057},
  year={2022}
}

@Article{Williams2009,
  author    = {S. Williams and A. Waterman and D. A. Patterson},
  journal   = {Commun. {ACM}},
  title     = {Roofline: an insightful visual performance model for multicore architectures},
  year      = {2009},
  number    = {4},
  pages     = {65--76},
  volume    = {52},
  bibsource = {dblp computer science bibliography, https://dblp.org},
  doi       = {10.1145/1498765.1498785},
}

@article{mohan2021analyzing,
  title={Analyzing and mitigating data stalls in DNN training},
  author={Mohan, Jayashree and Phanishayee, Amar and Raniwala, Ashish and Chidambaram, Vijay},
  journal={Proc. of the VLDB Endowment},
  volume={14},
  number={5},
  year={2021}
}

@inproceedings{gu2022fluid,
  title={Fluid: Dataset abstraction and elastic acceleration for cloud-native deep learning training jobs},
  author={Gu, Rong and Zhang, Kai and Xu, Zhihao and et al.},
  booktitle={2022 IEEE 38th Int. Conf. on Data Engineering},
  pages={2182--2195},
  year={2022},
  organization={IEEE}
}

@inproceedings{lee2021refurbish,
  title={Refurbish your training data: Reusing partially augmented samples for faster deep neural network training},
  author={Lee, Gyewon and Lee, Irene and Ha, Hyeonmin and et al.},
  booktitle={2021 USENIX Annual Technical Conf.},
  pages={537--550},
  year={2021}
}

@software{FastFlow,
    title        = {FastFlow},
    author       = {Taegeon, Um and Byungsoo, Oh and Byeongchan, Seo and et al.},
    year         = 2023,
    journal      = {GitHub repository},
    publisher    = {GitHub},
    howpublished = {https://github.com/SamsungLabs/
FastFlow}
}

@inproceedings{park2020trainbox,
  title={TrainBox: An extreme-scale neural network training server architecture by systematically balancing operations},
  author={Park, Pyeongsu and Jeong, Heetaek and Kim, Jangwoo},
  booktitle={53rd Annual IEEE/ACM Int. Symp. on Microarchitecture},
  pages={825--838},
  year={2020},
  organization={IEEE}
}

\begin{IEEEbiography}[{\includegraphics[width=1in,height=1.25in,clip,keepaspectratio]{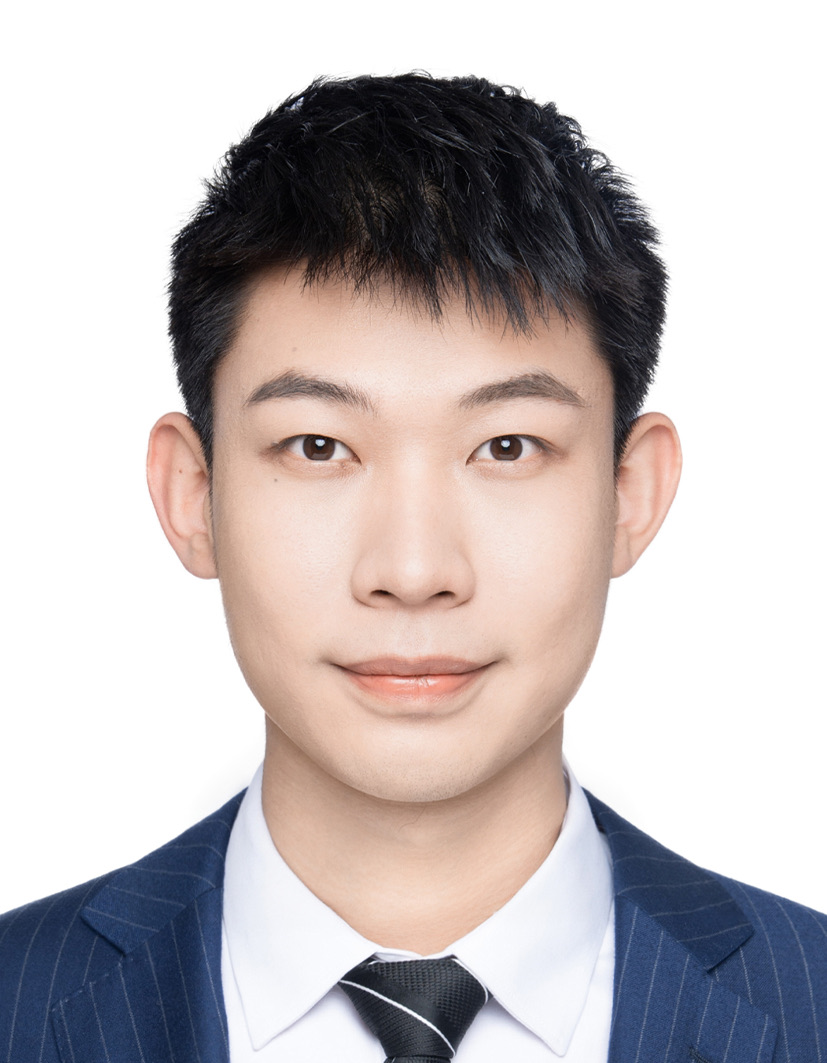}}]{Jia Wei}
received his Ph.D.degree in Computer Architecture from Xi’an Jiaotong University, China, in 2024. He is now a Postdoctoral Fellow at the Department of Computer Science and Technology of Tsinghua University, China. His research interests include AI System, computer architecture, highperformance computing, and deep learning.
\end{IEEEbiography}

\begin{IEEEbiography}[{\includegraphics[width=1in,height=1.25in,clip,keepaspectratio]{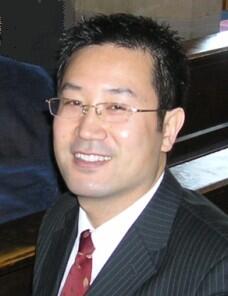}}]{Xingjun Zhang}
(Member, IEEE) received his Ph.D.degree in Computer Architecture from Xi’an Jiaotong University, China, in 2003. From Jan. 2004 to Dec. 2005, he was Postdoctoral Fellow at the Computer School of Beihang University, China. He is now a Full Professor and the Dean of the School of Computer Science \& Technology, Xi’an Jiaotong University. His research interests include high performance computing, big data storage system and machine learning acceleration.
\end{IEEEbiography}

\begin{IEEEbiography}[{\includegraphics[width=1in,height=1.25in,clip,keepaspectratio]{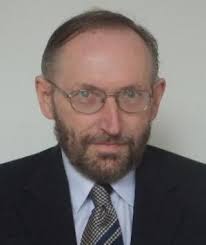}}]{Witold Pedrycz (Fellow, IEEE)}
received the M.Sc., Ph.D., and D.Sc. degrees from Silesian Technical University, Gliwice, Poland, in 1977, 1980, and 1984, respectively.
He is Professor with the Department of Electrical and Computer Engineering, University of Alberta, Edmonton, Canada. He has authored and coauthored numerous papers in these areas; the current h-index is 109 (Google Scholar) and 82 on the list top-h scientists for computer science and electronics.
\end{IEEEbiography}
\begin{IEEEbiography}[{\includegraphics[width=1in,height=1.25in,clip,keepaspectratio]{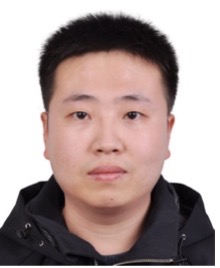}}]{Longxiang Wang}
received his Ph.D degree in Computer science from Xi’an Jiaotong University (XJTU), Xi’an, China. From 2017, he was a Lecturer in the Department of Computer Science \& Technology of Xi’an Jiao-tong University. His interests include high performance computing, storage system, and big data.
\end{IEEEbiography}

\begin{IEEEbiography}[{\includegraphics[width=1in,height=1.25in,clip,keepaspectratio]{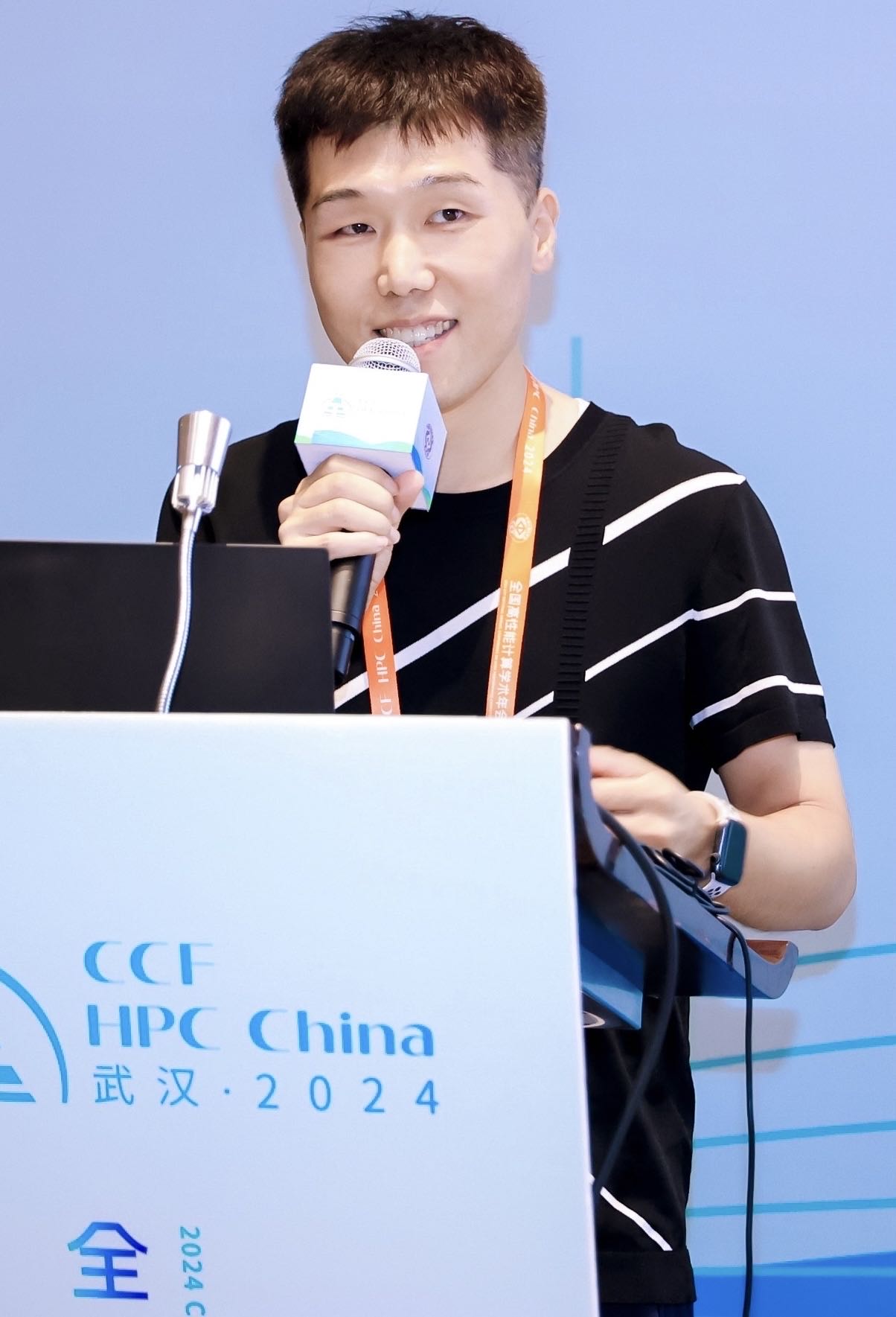}}]{Jie Zhao}
received his Ph.D degree in Mathematics from PSL Research University,Paris, France. From 2017, he was a Full Professor at the College of Computer Science and Electronic Engineering, Hunan University. His research interests focus on building intelligent software systems, with an emphasis on machine learning systems, polyhedral compilers, numerical program analysis, and high-level synthesis. 
\end{IEEEbiography}

\end{document}